\documentclass[%
 reprint,
 amsmath,amssymb,
 aps,
]{revtex4-1}

\usepackage{graphicx}
\usepackage{dcolumn}
\usepackage{bm}

\def\be{\begin{equation}}
\def\ee{\end{equation}}

\begin{document}

\preprint{APS/123-QED}

\title{Effects of strong magnetic fields and rotation on white dwarf structure}

\author{B. Franzon}
 \altaffiliation[]{franzon@fias.uni-frankfurt.de}
\author{ S. Schramm }%
 \email{schramm@fias.uni-frankfurt.de}
\affiliation{%
 Frankfurt Institute for Advanced Studies,
Ruth-Moufang - 1 60438, Frankfurt am Main,
Germany\\
}%




\date{\today}

\begin{abstract}
 In this work we compute  models for relativistic white dwarfs in the presence of strong magnetic fields.  These models  possibly contribute to super-luminous SNIa.
   With an assumed axi-symmetric and poloidal magnetic field, we study the possibility of existence of super-Chandrasekhar magnetized white dwarfs by solving numerically the Einstein-Maxwell equations, by means of a pseudo-spectral method.  We obtain a self-consistent rotating and non-rotating magnetized white dwarf models. According to our results, a maximum mass for a static magnetized white dwarf is 2.13 $\rm{M_{\odot}}$ in the Newtonian case and  2.09 $\rm{M_{\odot}}$ while taking into account general relativistic effects.
    Furthermore, we present results for rotating magnetized white dwarfs. The maximum magnetic field strength reached at the center of  white dwarfs  is of the order of $10^{15}\,$G in the static case, whereas for magnetized white dwarfs, rotating with the Keplerian angular velocity,  is of the order of $10^{14}\,$G. 
\end{abstract}

\pacs{95.30.Sf, 04.40.Dg, 97.10.Kc, 97.10.Ld}
\maketitle


\section{Introduction}
 White dwarfs (WDs) are stellar remnants of stars with masses of up to several solar masses. With a mass comparable to that of the Sun $\sim 2 \times 10^{33}\, \rm{g}$,  which is distributed in a volume comparable to that of the Earth,  the central mass density in these objects can reach values of about $10^{11} \, \rm{g/cm^{3}}$.   Together with neutron stars and black holes,  they are the endpoints of stellar evolutions and play a key role in astrophysics \cite{shapiro2008black, glendenning2012compact}.  

 The existence of white dwarfs was one of the major puzzles in astrophysics until Fowler \cite{fowler1926dense}, based on the quantum-statistical theory  developed by Fermi and Dirac  \cite{dirac1926theory, fermi1926sulla}, shows that white dwarfs are supported by the pressure of a degenerate electron gas. In addition,   Chandrasekhar in Ref. \cite{chandrasekhar1931maximum} included effects of special relativity in the degenerate electron gas theory and, as a result,  found that there is a limit in the stellar mass, above which degenerate white dwarfs are unstable. This critical mass is the so-called {\it Chandrasekhar limit} and is about 1.4 $\rm{M_{\odot}}$. 
 
White dwarfs are mostly composed of electron-degenerate matter.  The mass of the star is essentially due to the nuclei, whereas the main contribution to the pressure comes from the electrons.  For typical WDs,  thermonuclear reactions terminate at lighter nuclei, as Carbon, helium, or oxygen \cite{chandrasekhar1939, glendenning2012compact}.


Some white dwarfs are also associated with strong magnetic fields.  From observations, the surface magnetic field of these stars can reach values from $10^{6}\,$ G to  $10^{9}\,$ G  \cite{Terada:2007br, Reimers:1995ia, Schmidt:1995eh, Kemp:1970zz, putney1995three}.  However,  the internal magnetic field in magnetic stars is very poorly constrained by the observations and can be much stronger than in the surface. 
For example,  white dwarfs can have internal magnetic fields  as large as $10^{12-16}\,$ G according to Refs.~\cite{angel1978magnetic, shapiro2008black, Bera:2014wja}.   Moreover,  self-consistently highly magnetized neutron stars calculations have shown that neutron  stars can possess central magnetic fields as large as  $10^{18}\,$ G \cite{Bocquet:1995je, Chatterjee:2014qsa, cardall2001effects, Franzon:2015sya}.  Therefore,  understanding and estimating magnetic fields inside compact objects   occupy a key position in astrophysics. 

 Motivated by observations of a thermonuclear supernova that appears to be more luminous than expected  (e.g. SN 2003fg, SN 2006gz, SN 2007if,
SN 2009dc), it has been argued \cite{Scalzo:2010xd, Howell:2006vn, Hicken:2007ap, Yamanaka:2009dp, taubenberger2011high} that the progenitor of  such  super-novae should be a white dwarf with mass above the well-known Chandrasekhar limit, in other words, a  super-Chandrasekhar white dwarf. 


Progenitors with masses $\rm{M > 2.0\,\, \rm{M_{\odot}}}$ were considered in the literature as a result of mergers of two massive white dwarfs, or due to fast rotation \cite{Moll:2013mpa}.  In addition, super-Chandrasekhar white dwarfs were investigated in a strong magnetic field regime as in Refs.~\cite{Das:2012ai,Das:2013gd,Das:2014ssa}.  In the Newtonian framework,  models for white dwarfs endowed with a magnetic field and/or rotating were investigated in a series of papers \cite{adam1986models, ostriker1968rapidly, ostriker1968rapidly1, ostriker1969oscillations}. Recently, a study of differentially rotating and magnetized white dwarfs were performed within the ideal magnetohydrodynamic (GRMHD) regime \cite{Subramanian:2015sza} and they have found that differential rotation can increase the mass of magnetized white dwarfs up to 3.1 $\rm{M_{\odot}}$.

In Ref.~\cite{Das:2013gd}, it was obtained ultramagnetized white dwarfs with a maximum mass of 2.58 $\rm{M_{\odot}}$. The authors also found a magnetic field strength up to $10^{18}\,$G at the center of the star. Nonetheless, such approach violates not only macro physics aspects, as for example, the breaking of spherical symmetry due to the magnetic field, but also micro physics considerations, which are relevant for a self-consistent calculations of the structure of these objects  \cite{Coelho:2013bba, PhysRevD.88.081301, PhysRevD.91.028301}. Besides,  a self-consistently newtonian structure calculation of strongly magnetized white dwarfs shown that these stars exceed the traditional Chandrasekhar mass limit significantly $(\rm{M \sim 1.9\, M_{\odot}} )$ for a maximum field strength of the order of $\rm{10^{14}\,}$G  \cite{Bera:2014wja}.

In this work, we model static and rotating magnetized white dwarfs in a self-consistent way  by solving Einstein-Maxwell equations in the same way as done originally for neutron stars in Refs.~\cite{Bonazzola:1993zz, Bocquet:1995je}. We follow also our recent work on highly magnetized hybrid stars \cite{Franzon:2015sya}, where both general relativistic effects and the anisotropy of the energy momentum tensor caused by the magnetic field were taken into consideration to calculate the star structure. The presence of such a strong magnetic field can locally affect the microphysics  of the equation of state (EoS), as for example, due to Landau quantization.  As shown in Ref.~\cite{Bera:2014wja},  the Landau quantization does not affect the global properties of white dwarfs. However,  the authors solve the structure equation in a Newtonian form and without the magnetization term for the matter. Moreover, as we will see,  effects of general relativity can play an important role by determining the maximum mass of such highly magnetized white dwarfs.

Globally, the magnetic field can affect the structure of WDs since it contributes to the Lorenz force, which acts against gravity. In addition, it contributes also to the structure of the spacetime, since the magnetic field is now a source for the gravitational field through the Maxwell energy-momentum tensor. As we are interested in global effects that magnetic fields and the rotation can induce in WDs, we simplify the discussion assuming white dwarfs composed predominately by $^{12}C$ ($A/Z = 2$) in a electron background .

\section{Basic equations and formalism}
In this work, we consider rotating and non-rotating magnetized white dwarfs. The formalism used here was first applied to stationary neutron stars \cite{Bonazzola:1993zz, Bocquet:1995je, Chatterjee:2014qsa} and more recently in Ref.~\cite{Franzon:2015sya}.  Details of the gravitational equations,  numerical procedure and other properties of the equations can be found in the references cited above and in Ref.~\cite{gourgoulhon20123+}.  For the sake of completeness and for a better understanding of the reader, we show here some of the electromagnetic equations that, together with the gravitational equations, are solved numerically. The energy-momentum tensor of the system reads:
\be
T_{\alpha\beta} = (e+p)u_{\alpha}u_{\beta} + pg_{\alpha\beta} + \frac{1}{\mu_{0}} \left( F_{\alpha \mu} F^{\mu}_{\beta} - \frac{1}{4} F_{\mu\nu} F^{\mu\nu} \mathrm{g}_{\alpha\beta} \right),
\label{emt}
\ee
where $F_{\alpha\mu}$ denotes the antisymmetric Faraday tensor,  $e$ the energy density and $p$  the pressure as measured  by an observer ($\mathcal{O}_{1}$) co-moving with the fluid and whose  four-velocity is $u_{\alpha}$. The $g_{\alpha\beta}$ is the metric tensor. The first term in Eq.~\eqref{emt} is the  isotropic matter contribution to the energy momentum-tensor,  while the second term is the anisotropic electromagnetic field contribution.

The electromagnetic contribution (EM) to the energy-momentum tensor is obtained within the so-called 3+1 decomposition \cite{gourgoulhon20123+, Bonazzola:1993zz}. The energy density becomes:
\be
E^{(EM)} = \frac{1}{2\mu_{0}} ( E^i E_{i}  + B^i B_{i}      ),
\label{energydensity}
\ee
and the momentum-density flux can be written as:
\be
J^{(EM)}_{\phi} = \frac{1}{\mu_{0}} A^2(B^r E^\theta  - E^r B^\theta).
\label{momentumdensity}
\ee
The stress 3-tensor components are given by:
\be
S^{(EM) r}_{\;\;\; r} =  \frac{1}{2\mu_{0}} (E^\theta E_{\theta}  -    E^rE_{r} +  B^\theta B_{\theta} -  B^r B_{r}),
\label{stress1}
\ee
\be
S^{(EM) \theta}_{\;\;\; \theta} =  \frac{1}{2\mu_{0}} (E^r E_{r}  -    E^\theta E_{\theta} +  B^r B_{r}  -  B^\theta B_{\theta}),
\label{stress2}
\ee
\be
S^{(EM) \phi}_{\;\;\; \phi} =  \frac{1}{2\mu_{0}} ( E^i E_{i}   + B^i B_{i} ),
\label{stress3}
\ee
being the electric field components, as measured by the Eulerian observer $\mathcal{O}_{0}$,  written as \cite{lichnerowicz1967relativistic}:
\be
E_{\alpha} = \left( 0 , \frac{1}{N} \left[  \frac{\partial A_{t}}{\partial r} + N^{\phi} \frac{\partial A_{\phi}}{\partial r}\right ] , \frac{1}{N} \left[  \frac{\partial A_{t}}{\partial \theta} + N^{\phi} \frac{\partial A_{\phi}}{\partial \theta}\right ]   , 0 \right),
\ee
and the magnetic field  given by:
\be
\hspace{-2cm} B_{\alpha} = \left( 0 , \frac{1}{\Psi r^{2} \sin \theta} \frac{\partial A_{\phi}}{\partial \theta}, - \frac{1}{\Psi \sin \theta} \frac{\partial A_{\phi}}{\partial r} , 0  \right),
\ee
with $N^{\phi} (r, \theta)$ being the shift vector, $N (r, \theta)$ the lapse function and $\Psi$ a metric potential (for more details see Refs. \cite{Bonazzola:1993zz, Bocquet:1995je, Chatterjee:2014qsa}).  As in Ref.~\cite{Bonazzola:1993zz},  the equation of motion ($\nabla_{\mu}T^{\mu\nu}= 0$) reads:
\be
H \left(r, \theta \right) + \nu \left(r, \theta \right) -ln \Gamma \left( r, \theta \right) + M \left(r, \theta \right) = const,
\label{equationofmotion}
\ee
with  $H(r,\theta)$ being the logarithm of the dimensionless relativistic enthalpy per baryon:
\be
H:= ln \left( \frac{e+p}{m_{b}n_{b}c^{2}  } \right),
\label{enthalpy}
\ee
with $m_{b}$  the mean baryon mass $1.66\times10^{-27} \,\rm{kg}$,  $n_{b}$ the baryon number density. The second term in Eq.~\eqref{equationofmotion}  is defined as  $\nu = \nu(r,\theta):=ln (N)$ and  the Lorenz factor written as $\Gamma = (1 - U^{2})^{-\frac{1}{2}}$. The physical fluid velocity $U$  in the $\phi$ direction is defined  as:
\be
U = \frac{\Psi r\sin\theta}{N}(\Omega - N^\phi),
\ee
and the magnetic potential $M(r,\theta)$ associated to the Lorentz force is written as:
\be
M \left(r, \theta \right) = M \left( A_{\phi} \left(r, \theta \right) \right): = - \int^{0}_{A_{\phi}\left(r, \theta \right)} f\left(x\right) \mathrm{d}x,
\ee
with a current function $f(x)$ as defined in Ref.~\cite{Bonazzola:1993zz} (see eq. (5.29)). In this work, we use a current function $f(x)=f_{0}=const$, which is proportional to the intensity of the magnetic field, i.e., the higher the current function $f_{0}$, the higher is the magnetic field in the star. As shown in Ref.~\cite{Bocquet:1995je}, other choices are possible for $f(x)$, however, they do not alter the conclusions.

\section{Mass-radius diagram for static highly magnetized white dwarfs}
In this section we present the mass-radius (MR) diagram for static magnetized white dwarfs.  
The relation between the mass and the radius for non-magnetized white dwarfs  was first determined by Chandrasekhar~\cite{chandrasekhar1939}. Recently, studies on modified mass-radius relation of magnetic white dwarfs were proposed, for example, in Refs.~\cite{Das:2012ai, suh2000mass, Bera:2014wja}.  As we also found in this work, those authors shown that the mass of white dwarfs increase in presence of magnetic fields. 

\begin{figure}[!t]
\center\includegraphics[width=1.0\textwidth,angle=-90,scale=0.5]{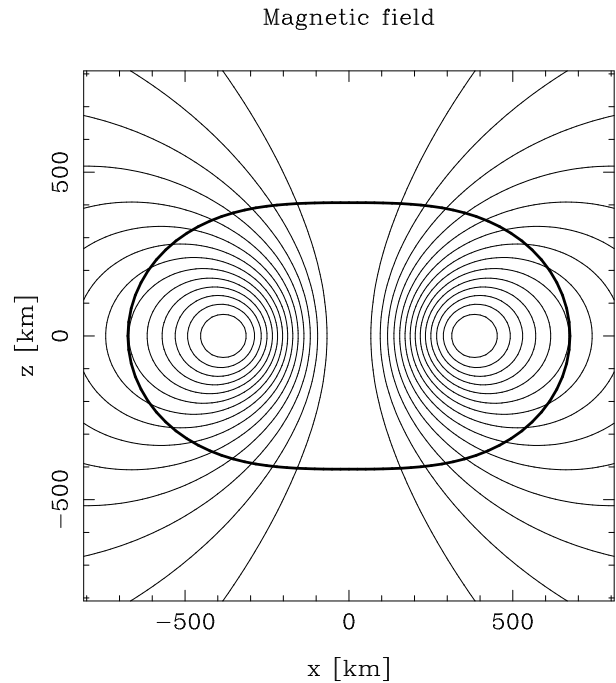}
\caption{Isocontours of the magnetic field strength in the $(x, z)$ plane, with a gravitational mass of 2.09 $\rm{M_{\odot}}$ and a magnetic dipole moment of $ 1.30\times 10^{34} \,\,\rm{Am^{2}}$. The ratio between the magnetic pressure and the matter pressure at the center of the star is about 1, and the magnetic field at center reaches 1.03$\times10^{15}\,$G.}
\label{bfield_static}
\end{figure}

In Fig.~\ref{bfield_static}, we show the isocontours in the $(x, z)$ plane of the poloidal magnetic field lines for a static star with central enthalpy  of $H_{c} = 0.0063\,c^{2}$. As we will see in Fig.~\ref{wd_static},  this value of the enthalpy gives us the maximum gravitational mass of relativistic, static and magnetized white dwarfs achieved with the code, namely,  2.09 $\rm{M_{\odot}}$, which corresponds to a central mass density of 2.79$\times 10^{10}\,\rm{g/cm^{3}}$. It is known that at sufficiently high densities reactions as inverse $\beta$-decay or pycnonuclear fusion can take place in the interior of white dwarfs \cite{PhysRevD.88.081301, PhysRevD.92.023008}. As estimated in Ref.~\cite{PhysRevD.92.023008}, for the maximum mass configuration obtained in this work, i.e., for a central  magnetic field of $1.03\times10^{15}\,$G, the onset of electron capture by carbon-12 nuclei was found to be about $4.2\times10^{10}\,\rm{g/cm^{3}}$ with electron-ion interactions, and $3.9\times10^{10}\,\rm{g/cm^{3}}$ without electron-ion interactions. Therefore, in our calculation, the maximum mass density reached by the most massive and non-rotating magnetized white dwarf lies below the threshold density for the onset of electron captures by carbon-12 nuclei as calculated  in Ref.~\cite{PhysRevD.92.023008}.   

\begin{figure}[!t]
\center
\includegraphics[width=1.0\textwidth,angle=-90,scale=0.5]{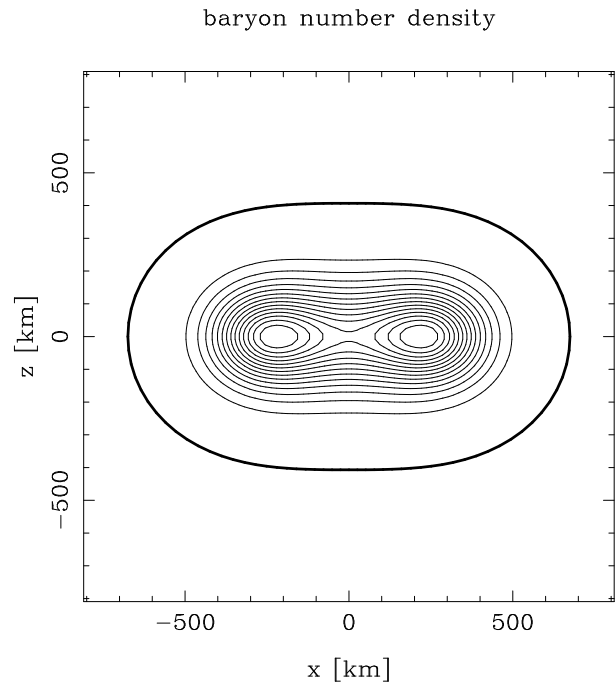}\caption{Isocontours of the baryon number density in the $(x, z)$ plane for the same star as shown in Fig.~\ref{bfield_static}. The central baryon density for this model is  $\rm{1.679 \times 10^{-5} \,\, f
m^{-3} \,\,(2.79 \times 10^{10}\,\, g/cm^{3})}$.}
\label{nb}
\end{figure}

\begin{figure}[!t]
\center
\includegraphics[width=0.7\textwidth,angle=-90,scale=0.5]{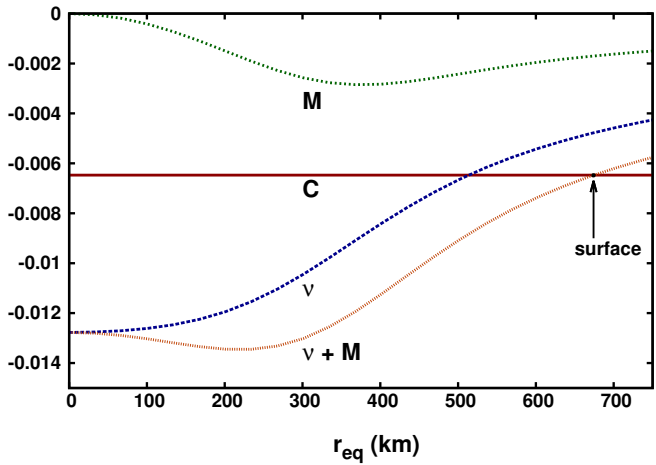}
\caption{Behaviour of the different terms of the equation of motion as a function of the equatorial coordinate radius for the same star as shown in Fig. \ref{bfield_static}.}
\label{eqm}
\end{figure}

\begin{figure}[!t]
\center
\includegraphics[width=0.7\textwidth,angle=-90,scale=0.5]{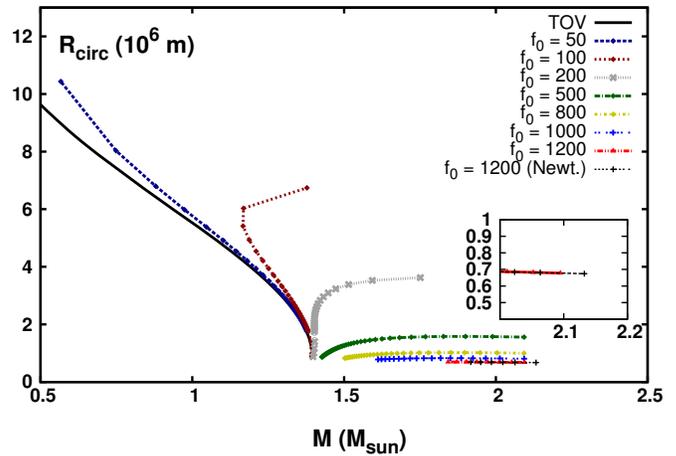}
\caption{ Mass-radius diagram for magnetized white dwarfs. Different curves represent different values of the current function $f_{0}$. We also compare the maximum white dwarf mass obtained in the Newtonian case (in black) and in the relativistic one (in red) for the maximum electric current value for which  numerical convergence is achieved. This diagram is quite similar to the MR diagram calculated in Ref.~\cite{Bera:2014wja}. However, those authors have evaluated only Newtonian white dwarfs. All curves in this figure were calculated for a ratio between the magnetic and the matter pressure less than or quite close to 1 at the center of the star.}
\label{wd_static}
\end{figure}

 In Fig.~\ref{nb},  we show the mass density distribution for the same star as shown in Fig.~\ref{bfield_static}.  As expected, the mass density is not spherically distributed and the maximum mass density is not at the center of the star. In this case,  the central magnetic field  reaches a value of 1.03$\times 10^{15}\,$G,  whereas the surface magnetic field was found to be 2.02 $\times 10^{14}\,$G.   As a result, one sees that the Lorentz force exerted by the magnetic field  breaks the spherical symmetry of the star considerably and behaves as a centrifugal force,  which pushes the matter off-center.
In order to understand better this aspect, we can use the equation of motion (\ref{equationofmotion}) for the static case $\Gamma = 0 $:
\be
 H(r,\theta) + \nu (r, \theta) +  M(r,\theta) =  C,
\label{bernuli}
\ee
and then plot these quantities in the equatorial plane as shown in Fig.~\ref{eqm}. This same analysis was presented for magnetized neutron stars in Ref.~\cite{cardall2001effects}. The constant $C$ can be calculated at every point in the star.  We have chosen the center, since the central values of the magnetic potential $M(r,\theta)$ is zero and the central enthalpy $H_{c}$ is our input to construct the models. The Lorentz force is the derivative of the magnetic potential $M(r, \theta)$ in the equation Eq.~\eqref{bernuli} and reaches its maximum value off-center ($r_{eq}\sim$ 350 km, see Fig.~\ref{eqm}).
As already discussed  in Ref.~\cite{cardall2001effects}, the  direction of the magnetic forces in the equatorial plane depends on the current distribution inside the star. In addition, the magnetic field changes its direction in the equatorial plane  and, therefore, the Lorenz force reverses the direction inside the star. In our case,   this can be seen from  the qualitatively change in behaviour of the function $M(r, \theta)$ around $\rm{r_{eq}} \sim$ 350 km in Fig.~\ref{eqm}.

In order to compare  our results with those in the literature, we compute the mass-radius diagram for magnetized white dwarfs as shown in Fig.~\ref{wd_static}.
As pointed out in Ref.~\cite{paret2015maximum},  a fully consistent equilibrium configuration for magnetized white dwarfs is still lacking.  Those same authors have used a relativistic framework with cylindrical coordinates and the splitting of the pressure into parallel and perpendicular components to show that the maximum magnetic field inside these objects can not exceed $\rm{1.5 \times 10^{13}}\,$ G. Their solutions also indicate that it is not possible to have stable magnetized WDs with super-Chandrasekhar masses.  The maximum magnetic field strength obtained in Ref.~\cite{paret2015maximum} is less than the value obtained by a self-consistent  solution of Newtonian white dwarfs as presented in Ref.~\cite{Bera:2014wja},  and orders of magnitude less than predicted in Ref.~\cite{Das:2012ai}. However, as shown in Refs.~\cite{Coelho:2013bba, PhysRevD.91.028301, PhysRevD.88.081301}, the work  \cite{Das:2012ai} has been criticized above all because their calculation violate both macro/micro physics properties essential for the stability of these objects.

 In Fig.~\ref{wd_static}, the maximum white dwarf masses are obtained when the ratio between the magnetic and the matter pressure at the center of the star is less than or about 1.  For such a strong magnetic field,  the magnetic force has pushed the matter off-center and a topological change to a toroidal configuration can take place \cite{cardall2001effects}. This gives a limit for the magnetic field strength that can be computed within this approach, since our current numerical tools do not enable us to handle
toroidal configuration. 

The maximum masses of white dwarfs increase with the magnetic field.  In our calculation, we found a relativistic white dwarf mass of 2.09 $\rm{M_{\odot}}$ for almost the same magnetic field strength at the center,   B $\sim \rm{10^{14}}\,$ G,  as in Ref.~\cite{Bera:2014wja}.   The same authors presented configurations with magnetic fields up to $\rm{10^{16}\, G }$, which we have not found in our calculations. In the Newtonian case, we found a mass of 2.13 $\rm{M_{\odot}}$  for the most massive magnetized white dwarf. In both cases, the masses are well above the Chandrasekhar limit of 1.4 $\rm{M_{\odot}}$.
 \section{Rotating magnetized white dwarfs}
Aside magnetic fields, rotation is a crucial observable in stellar astrophysics. In the case of neutron stars, for example, the magnetic dipole model states a direct relation between the rotation period/period derivative with the polar magnetic field of the star. For white dwarfs the interest in understanding these objects has been increasing with the years. Some observed white dwarfs rotate with periods of days or even years. One of the fastest observed WD possesses a spin period of $13.2\,s$ \cite{Mereghetti:2010id}, a value similar to those observed in Soft Gamma Repeaters (SGRs) and  Anomalous X-ray pulsars (AXPs), known as magnetars \cite{Duncan:1992hi, Thompson:1993hn}.  A relation between white dwarfs and magnetars was addressed  in Ref.~\cite{Malheiro:2015yda},  where the authors speculated that SGRs and AXPs with low magnetic field on the surface might be rotating magnetized white dwarfs.

Rigidly rotating non-magnetized white dwarfs were already studied long time ago in the Newtoninan framework as in Refs.~\cite{krishan1963limiting, anand1965chandrasekhar, james1964structure, roxburgh1966structure, monaghan1966structure, geroyannis1989models} . In addition, the structure of rapidly rotating white dwarfs was performed  in general relativity 
as in Ref.~\cite{arutyunyan1971rotating}, and more recently in Ref.~\cite{ boshkayev2013general}, where the authors used the Hartle's formalism \cite{hartle1967slowly} to solve the Einstein equations.  
 From the standpoint of rotation, it is clear that all rotating stars have to satisfy the mass-shedding, or Keplerian limit, as a condition of stability.  This limit is reached when the centrifugal force due to the rotation does not balance gravity anymore and the star starts to lose particles from the equator, defining an upper limit to the angular velocity of uniformly rotating stars.

In the same way that rotation provides a natural limit for the stability of stars, in the case of WDs, there are also microphysics aspects, as for example, the inverse $\beta-$decay and pycnonuclear fusion reactions  \cite{PhysRevD.88.081301, PhysRevD.92.023008, Coelho:2013bba},  that need to be taken into account for a complete and self-consistent description of these stars.  In this paper,  however, we restrict ourselves to the study of the combined effect of rotation and magnetic fields on the global structure of WDs and we do not address here the microphysics aspects.
\begin{figure}[!t]
\center
\includegraphics[width=0.7\textwidth,angle=-90,scale=0.5]{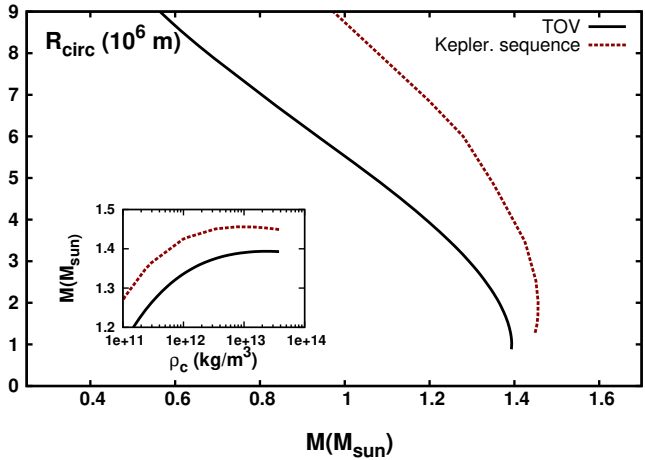}
\caption{Mass-radius diagram for static and rotating white dwarfs.  The TOV solution is shown in black, and in red (dashed) we show the Keplerian sequence for rotating WDs. On the bottom left-hand corner we show the gravitational mass as a function of the central density for the same sequence of stars.  }
\label{wd_rot}
\end{figure}

In Fig.~\ref{wd_rot},  we show the Tolman-Oppenheimer-Volkoff (TOV) solution for the structure of a spherically symmetric white dwarf and the mass-shedding frequency limit. The centrifugal force exerted by the rotation acts against gravity, which allows the star to support higher masses compared to the static case. In the first place, by comparison of Fig.~\ref{wd_static} and Fig.~\ref{wd_rot}, one sees that magnetic fields are more efficient than rotation in increasing the maximum mass of stars. The maximum mass obtained for a relativistic and magnetic white dwarf is 2.09 $\rm{M_{\odot}}$, whereas the maximum mass achieved by rotation is $\sim$1.45$\,\rm{M_{\odot}}$.  

\begin{figure}[!t]
\center
\includegraphics[width=0.7\textwidth,angle=-90,scale=0.5]{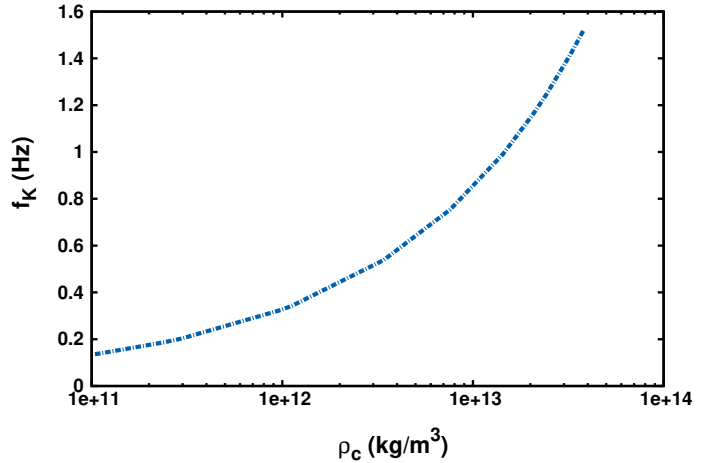}
\caption{ Keplerian frequency as a function of the central baryon density for the sequence as shown in Fig.~\ref{wd_rot}. The maximum frequency reached by a non-magnetized and uniformilly rotating white dwarf is 1.52 $\rm{Hz}$.}
\label{wd_freq}
\end{figure}

In Fig.~\ref{wd_freq},  we show the relation between the Keplerian frequency ($f_{K}$) and the central density of the star. The higher angular velocity, the higher centrifugal forces, which pushes the matter outward, and therefore, acting against gravity. As a result, the stars are allowed to have more mass and, then, increasing the central density.  This is possible, because the centrifugal forces due to rotation ($f_{c} \propto r\Omega^2$) have much more effect on the outer layers of the star. On the other hand, for  non-rotating magnetized stars, the Lorenz force acts mainly in the inner layers of the star, reducing, and not increasing, the central densities in these objects as shown in Ref.~\cite{Franzon:2015sya}. 

\begin{figure}[!t]
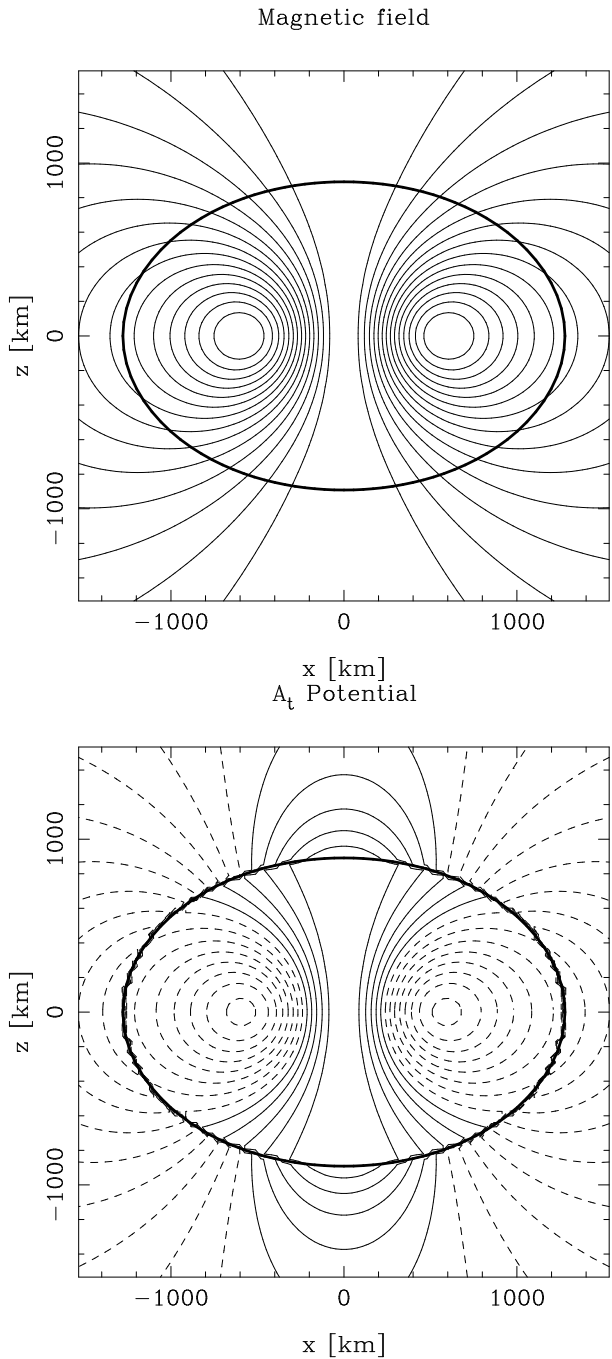

\begin{center}
\includegraphics[width=1\textwidth,angle=-90,scale=0.5]{bfield.eps} \quad
        \includegraphics[width=1\textwidth,angle=-90,scale=0.5]{pgplot.eps}
\caption{Magnetic field lines (up) and electromagnetic potential lines (down) for a star with $H_{c} = 0.005\,c^{2}$ and graviational mass of $\sim$1.57 $\rm{M_{\odot}}$, with a Keplerian frequency of 1.13 $\rm{Hz}$. The dipole magnetic moment reaches 6.93$\times10^{33}\,\rm{Am^{2}}$ and the magnetic field intensity at the center of this white dwarf is 1.87 $\times 10^{14}\,$G.
} 
\label{bplusrotation}
\end{center}
\end{figure}

Henceforth we will investigate the role played by the magnetic field in uniformly rotating white dwarfs .  For a star  with central enthalpy of $H_{c} = 0.005\,c^{2}$, whose mass is close to the maximum mass in the static case,  we present results of three different calculations: $A)$ static and non$-$magnetized; $B)$ rotating (with the Keplerian frequency) and non-magnetized and $C)$ rotating (with the Keplerian frequency) and magnetized. 
The cases $A)$ and $B)$ are presented in the mass-radius diagram in Fig. 4 and \ref{wd_rot}.  For the case $B)$, the star rotates with its Keplerian frequency of 0.99 $\rm{Hz}$.  In addition, to construct the models $C)$, we turn on the magnetic field until that the limit of numerical convergence is reached. The resulting magnetic field lines and the electric isopotential lines $A_{t} = $ const  are depicted in Fig.~\ref{bplusrotation}.

\begin{figure}[!t]
\begin{center}
\includegraphics[width=0.7\textwidth,angle=-90,scale=0.5]{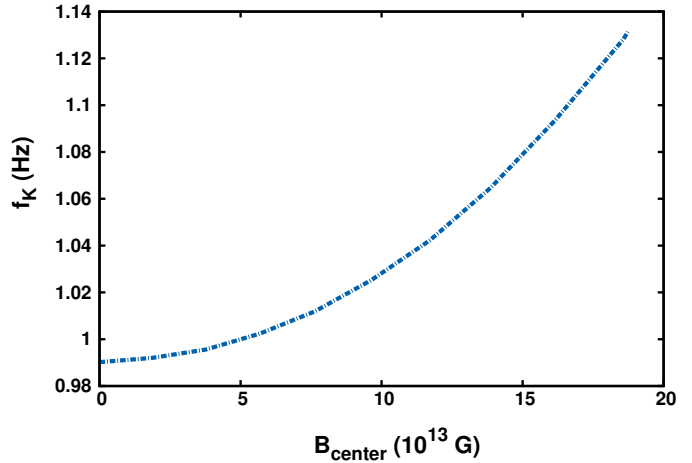}
\caption{Keplerian frequency as function of the central magnetic field for non-magnetized (case $B$, i.e., $f_{K}$ = 0.99 $\rm{Hz}$)  and magnetized  (case $C$, i.e., $f_{K}$ = 1.13 $\rm{Hz}$) white dwarfs. }
\label{fb}
\end{center}
\end{figure}

\begin{figure}[!t]\center\includegraphics[width=0.7\textwidth,angle=-90,scale=0.5]{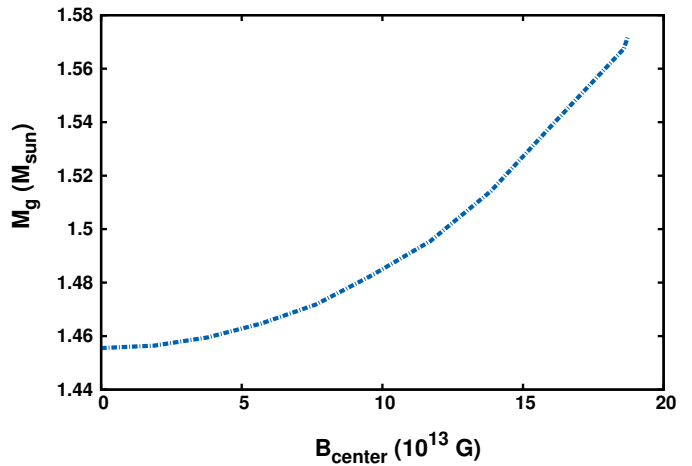}\caption{Gravitational mass as a function of the central magnetic field for a star with central enthalpy $H_{c} = 0.005\,c^{2}$ for the cases $B)$ and $C)$, which are the same stars as shown in Fig.~\ref{fb}.}
\label{mb}
\end{figure}

In order to study how the Keplerian frequency changes with the magnetic field, we perform a calculation for different current functions $f=f_{0}=const$, from zero (case $B)$ to the maximum value of the magnetic field as shown in Fig.~\ref{bplusrotation}.  As a result,  the Keplerian frequency  increases with the magnetic field as shown in Fig.~\ref{fb}.   In this way, equilibrium configurations are obtained for higher centrifugal forces and, therefore,  if the star can rotate faster, it can support higher masses, which explains the behaviour observed in Fig.~\ref{mb}. 

According to Figure \ref{wd_freq}, non-magnetized white dwarfs can reach a maximum Keplerian frequency of 1.52 $\rm{Hz}$. However, in the magnetic case (Fig.~\ref{fb}),  the maximum Keplerian frequency is reduced to 1.13 $\rm{Hz}$,  which corresponds to a white dwarf with gravitational mass of $\sim$ 1.57 $\rm{M_{\odot}}$ and a central magnetic field of 1.87 $\times 10^{14}\,$G.

\section{Conclusions}

We computed perfect-fluid magnetized white dwarfs in general relativity by solving the coupled Einstein-Maxwell equations. We have applied a formalism that was developed for neutron stars to rotating magnetized white dwarfs. In our case,  the equilibrium solutions are axisymmetric and stationary,  with white dwarfs endowed with a strong poloidal magnetic field.   
  
The observation of super-luminous Ia supernovae suggests that their progenitors are super-Chandrasekhar white dwarfs, whose masses are higher than 1.4 $\rm{M_{\odot}}$. The increasing in the mass may be a result from ultra-strong magnetic field inside the white dwarfs. The relevance of magnetic fields  in enhancing the maximum mass of a white dwarf were studied and the results were  obtained in a  fully relativistic framework. We have shown that white dwarfs masses  increase up to 2.09 $\rm{M_{\odot}}$ for a maximum magnetic field strength of $\sim10^{15}\,$G  in the stellar center. Thus, magnetic fields can, potentially, be responsible for super massive white dwarfs.

The structure of relativistic, axisymmetric and uniformly rotating magnetized white dwarfs were investigated self-consistently and all effects of electromagnetic field on the star equilibrium were taken into account.  As usual, non-magnetized configurations at the mass shedding limit support higher masses as their static counterparts.  However, we have seen that the magnetic field is much more efficient in increasing the mass of WDs than rotation.  For example, the Keplerian sequence has a maximum mass of $\sim$ 1.45 $\rm{M_{\odot}}$, whereas for the maximum magnetic field configuration achieved in this calculation, the maximum mass of relativistic WDs is 2.09 $\rm{M_{\odot}}$, i.e.,  33 $\%$ larger than the Chandrasekhar limit, and 2.13 $\rm{M_{\odot}}$  in the Newtonian framework.
  
 We have also shown the increasing of the Keplerian frequency ($f_{K}$) with the magnetic field.  The higher the magnetic field, the higher the Lorenz force, which, in turn,  helps the star to support more mass than in the non-magnetized case. As a result, if these stars  are more massive, they can rotate faster (see Fig.~\ref{fb} and Fig.~\ref{mb}). In this case,  the maximum mass obtained for  rotating magnetized white dwarfs is of $\sim$1.57 $\rm{M_{\odot}}$, with a Keplerian frequence of  1.13 $\rm{Hz}$. 
   
It is to be noted that purely poloidal or purely toroidal magnetic field configurations undergo intrinsic instabilities as suggested years ago in Refs.~\cite{markey1973adiabatic, tayler1973adiabatic, wright1973pinch, flowers1977evolution}. The nature of this instability was confirmed both in Newtonian numerical simulations as in Refs.~\cite{lander2012there, braithwaite2006stability, braithwaite2006stable, braithwaite2007stability} and in general relativity framework as in Refs.~\cite{ciolfi2013twisted, lasky2011hydromagnetic, marchant2011revisiting,  mitchell2015instability}.  
Analytical and numerical calculations have also shown that stable equilibrium configurations are obtained for magnetic fields composed not only by a poloidal component,  which extends throughout the  star and to the exterior, but also  a toroidal one, which is confined inside the star ~\cite{armaza2015magnetic, prendergast1956equilibrium, braithwaite2004fossil, braithwaite2006stable, akgun2013stability}. In addition, the magnetic field might decay due to ohmic effects and, therefore, changing its strength and distribution in the star \cite{goldreich1992magnetic}.   Although we model magnetized white dwarfs with a purely poloidal magnetic field components,  which are not the most general case,  we have shown, in a fully general relativity way,  that these stars can considerably increase their masses  due to magnetic field effects and, therefore, contributing to super-luminous  SNIa. In a future work, in order to have  a more complete description of the stars presented here, we intend to take into account magnetic field configurations with both poloidal and toroidal components, a B-field dependent equation of state and effects of the anomalous magnetic moment, too. 
\begin{acknowledgements}
The authors thank to  Nicolas Chamel for fruitful comments on the onset of the electron capture by cabon-12 nuclei and  the referee for the valuable comments and suggestions. B. Franzon acknowledges support from CNPq/Brazil, DAAD and HGS-HIRe for FAIR.  S. Schramm acknowledges support from the HIC for FAIR LOEWE program. 
\end{acknowledgements}

\nocite{*}

\bibstyle{apsrev4-1}
\bibliography{biblio}


@book{glendenning2012compact,
  title={Compact stars: Nuclear physics, particle physics and general relativity},
  author={Glendenning, Norman K},
  year={2012},
  publisher={Springer Science \& Business Media}
}

@article{Kemp:1970zz,
      author         = "Kemp, James C. and Swedlund, John B. and Landstreet, J.
                        D. and Angel, J. R. P.",
      title          = "{Discovery of Circularly Polarized Light from a White
                        Dwarf}",
      journal        = "Astrophys. J.",
      volume         = "161",
      year           = "1970",
      pages          = "L77-L79",
      doi            = "10.1086/180574",
      SLACcitation   = "
}

@article{Schmidt:1995eh,
      author         = "Schmidt, Gary D. and Smith, Paul S.",
      title          = "{A Search for magnetic fields among DA white dwarfs}",
      journal        = "Astrophys. J.",
      volume         = "448",
      year           = "1995",
      pages          = "305",
      doi            = "10.1086/175962",
      reportNumber   = "STEWARD-1250",
      SLACcitation   = "
}

@article{Reimers:1995ia,
      author         = "Reimers, D. and Jordan, S. and Koester, D. and Bade, N.
                        and Kohler, T. and Wisotzki, L.",
      title          = "{Discovery of four white dwarfs with strong magnetic
                        fields by the Hamburg / ESO survey}",
      journal        = "Astron. Astrophys.",
      volume         = "311",
      year           = "1996",
      pages          = "572-578",
      eprint         = "astro-ph/9604104",
      archivePrefix  = "arXiv",
      primaryClass   = "astro-ph",
      reportNumber   = "JORDAN-18-4-96",
      SLACcitation   = "
}



@article{Terada:2007br,
      author         = "Terada, Yukikatsu and Hayashi, Takayuki and Ishida,
                        Manabu and Mukai, Koji and Dotani, Tadayas u and Okada,
                        Shunsaku and Nakamura, Ryoko and Naik, Sachindra and
                        Bamba, Aya and Makishima, Kazuo",
      title          = "{Suzaku Discovery of Non-thermal X-ray Emission from the
                        Rotating Magnetized White Dwarf, AE Aquarii}",
      journal        = "Publ. Astron. Soc. Jap.",
      volume         = "60",
      year           = "2008",
      pages          = "387",
      doi            = "10.1093/pasj/60.2.387",
      eprint         = "0711.2716",
      archivePrefix  = "arXiv",
      primaryClass   = "astro-ph",
      SLACcitation   = "
}


@article{Terada:2007br,
      author         = "Terada, Yukikatsu and Hayashi, Takayuki and Ishida,
                        Manabu and Mukai, Koji and Dotani, Tadayas u and Okada,
                        Shunsaku and Nakamura, Ryoko and Naik, Sachindra and
                        Bamba, Aya and Makishima, Kazuo",
      title          = "{Suzaku Discovery of Non-thermal X-ray Emission from the
                        Rotating Magnetized White Dwarf, AE Aquarii}",
      journal        = "Publ. Astron. Soc. Jap.",
      volume         = "60",
      year           = "2008",
      pages          = "387",
      doi            = "10.1093/pasj/60.2.387",
      eprint         = "0711.2716",
      archivePrefix  = "arXiv",
      primaryClass   = "astro-ph",
      SLACcitation   = "
}




@article{Borra:1982jk,
      author         = "Borra, E. F. and Landstreet, J. D. and Mestel, L.",
      title          = "{Magnetic stars}",
      journal        = "Ann. Rev. Astron. Astrophys.",
      volume         = "20",
      year           = "1982",
      pages          = "191-220",
      doi            = "10.1146/annurev.aa.20.090182.001203",
      SLACcitation   = "
}

@article{Scalzo:2010xd,
      author         = "Scalzo, R. A. and others",
      title          = "{Nearby Supernova Factory Observations of SN 2007if:
                        First Total Mass Measurement of a Super-Chandrasekhar-Mass
                        Progenitor}",
      journal        = "Astrophys. J.",
      volume         = "713",
      year           = "2010",
      pages          = "1073-1094",
      doi            = "10.1088/0004-637X/713/2/1073",
      eprint         = "1003.2217",
      archivePrefix  = "arXiv",
      primaryClass   = "astro-ph.CO",
      SLACcitation   = "
}

@article{Howell:2006vn,
      author         = "Howell, D. Andrew and others",
      title          = "{The type Ia supernova SNLS-03D3bb from a
                        super-Chandrasekhar-mass white dwarf star}",
      collaboration  = "SNLS",
      journal        = "Nature",
      volume         = "443",
      year           = "2006",
      pages          = "308",
      doi            = "10.1038/nature05103",
      eprint         = "astro-ph/0609616",
      archivePrefix  = "arXiv",
      primaryClass   = "astro-ph",
      SLACcitation   = "
}




@article{Hicken:2007ap,
      author         = "Hicken, M. and Garnavich, P. M. and Prieto, J. L. and
                        Blondin, S. and DePoy, D. L. and Kirshner, R. P. and
                        Parrent, J.",
      title          = "{The Luminous and Carbon-Rich Supernova 2006gz: A Double
                        Degenerate Merger?}",
      journal        = "Astrophys. J.",
      volume         = "669",
      year           = "2007",
      pages          = "L17-L20",
      doi            = "10.1086/523301",
      eprint         = "0709.1501",
      archivePrefix  = "arXiv",
      primaryClass   = "astro-ph",
      SLACcitation   = "
}



@article{Yamanaka:2009dp,
      author         = "Yamanaka, M. and others",
      title          = "{Early phase observations of extremely luminous Type Ia
                        Supernova 2009dc}",
      journal        = "Astrophys. J.",
      volume         = "707",
      year           = "2009",
      pages          = "L118-L122",
      doi            = "10.1088/0004-637X/707/2/L118",
      eprint         = "0908.2059",
      archivePrefix  = "arXiv",
      primaryClass   = "astro-ph.HE",
      SLACcitation   = "
}

@article{chandrasekhar1931maximum,
  title={The maximum mass of ideal white dwarfs},
  author={Chandrasekhar, Subrahmanyan},
  journal={The Astrophysical Journal},
  volume={74},
  pages={81},
  year={1931}
}




@article{Moll:2013mpa,
      author         = "Moll, Rainer and Raskin, Cody and Kasen, Daniel and
                        Woosley, Stan",
      title          = "{Type Ia Supernovae from Merging White Dwarfs. I. Prompt
                        Detonations}",
      journal        = "Astrophys. J.",
      volume         = "785",
      year           = "2014",
      pages          = "105",
      doi            = "10.1088/0004-637X/785/2/105",
      eprint         = "1311.5008",
      archivePrefix  = "arXiv",
      primaryClass   = "astro-ph.HE",
      SLACcitation   = "
}

@article{Das:2012ai,
      author         = "Das, Upasana and Mukhopadhyay, Banibrata",
      title          = "{Strongly magnetized cold electron degenerate gas:
                        Mass-radius relation of the magnetized white dwarf}",
      journal        = "Phys. Rev.",
      volume         = "D86",
      year           = "2012",
      pages          = "042001",
      doi            = "10.1103/PhysRevD.86.042001",
      eprint         = "1204.1262",
      archivePrefix  = "arXiv",
      primaryClass   = "astro-ph.HE",
      SLACcitation   = "
}

@article{Das:2013gd,
      author         = "Das, Upasana and Mukhopadhyay, Banibrata",
      title          = "{New mass limit for white dwarfs: super-Chandrasekhar
                        type Ia supernova as a new standard candle}",
      journal        = "Phys.Rev.Lett.",
      number         = "7",
      volume         = "110",
      pages          = "071102",
      doi            = "10.1103/PhysRevLett.110.071102",
      year           = "2013",
      eprint         = "1301.5965",
      archivePrefix  = "arXiv",
      primaryClass   = "astro-ph.SR",
      SLACcitation   = "
}

@article{Das:2014ssa,
      author         = "Das, Upasana and Mukhopadhyay, Banibrata",
      title          = "{Maximum mass of stable magnetized highly
                        super-Chandrasekhar white dwarfs: stable solutions with
                        varying magnetic fields}",
      journal        = "JCAP",
      volume         = "1406",
      year           = "2014",
      pages          = "050",
      doi            = "10.1088/1475-7516/2014/06/050",
      eprint         = "1404.7627",
      archivePrefix  = "arXiv",
      primaryClass   = "astro-ph.SR",
      SLACcitation   = "
}


@article{Subramanian:2015sza,
      author         = "Subramanian, Sathyawageeswar and Mukhopadhyay, Banibrata",
      title          = "{GRMHD formulation of highly super-Chandrasekhar rotating
                        magnetised white dwarfs: Stable configurations of
                        non-spherical white dwarfs}",
      year           = "2015",
      eprint         = "1507.01606",
      archivePrefix  = "arXiv",
      primaryClass   = "astro-ph.SR",
      SLACcitation   = "
}



@article{Coelho:2013bba,
      author         = "Coelho, J.G. and Marinho, R.M. and Malheiro, M. and
                        Negreiros, R. and Rueda, J.A. and others",
      title          = "{Dynamical instability of white dwarfs and breaking of
                        spherical symmetry under the presence of extreme magnetic
                        fields}",
      journal        = "Astrophys.J.",
      number         = "1",
      volume         = "794",
      pages          = "86",
      doi            = "10.1088/0004-637X/794/1/86",
      year           = "2014",
      eprint         = "1306.4658",
      archivePrefix  = "arXiv",
      primaryClass   = "astro-ph.SR",
      SLACcitation   = "
}


@article{Bera:2014wja,
      author         = "Bera, Prasanta and Bhattacharya, Dipankar",
      title          = "{MassÐradius relation of strongly magnetized white
                        dwarfs: nearly independent of Landau quantization}",
      journal        = "Mon. Not. Roy. Astron. Soc.",
      volume         = "445",
      year           = "2014",
      number         = "4",
      pages          = "3951-3958",
      doi            = "10.1093/mnras/stu2014",
      eprint         = "1405.2282",
      archivePrefix  = "arXiv",
      primaryClass   = "astro-ph.SR",
      SLACcitation   = "
}

@article{Bonazzola:1993zz,
      author         = "Bonazzola, S. and Gourgoulhon, E. and Salgado, M. and
                        Marck, J. A. t Axisymmetric rotating relativistic bodies:
                        A new numerical approach for 'exact' solutions",
      title          = "{Axisymmetric rotating relativistic bodies: A new
                        numerical approach for 'exact' solutions}",
      journal        = "Astron. Astrophys.",
      volume         = "278",
      year           = "1993",
      pages          = "421-443",
      SLACcitation   = "
}


@article{Bocquet:1995je,
      author         = "Bocquet, M. and Bonazzola, S. and Gourgoulhon, E. and
                        Novak, J.",
      title          = "{Rotating neutron star models with magnetic field}",
      journal        = "Astron. Astrophys.",
      volume         = "301",
      year           = "1995",
      pages          = "757",
      eprint         = "gr-qc/9503044",
      archivePrefix  = "arXiv",
      primaryClass   = "gr-qc",
      SLACcitation   = "
}



@article{Chatterjee:2014qsa,
      author         = "Chatterjee, Debarati and Elghozi, Thomas and Novak,
                        Jerome and Oertel, Micaela",
      title          = "{Consistent neutron star models with magnetic field
                        dependent equations of state}",
      journal        = "Mon. Not. Roy. Astron. Soc.",
      volume         = "447",
      year           = "2015",
      pages          = "3785",
      doi            = "10.1093/mnras/stu2706",
      eprint         = "1410.6332",
      archivePrefix  = "arXiv",
      primaryClass   = "astro-ph.HE",
      SLACcitation   = "
}



@book{gourgoulhon20123+,
  title={3+ 1 formalism in general relativity: bases of numerical relativity},
  author={Gourgoulhon, Eric},
  volume={846},
  year={2012},
  publisher={Springer Science \& Business Media}
}

@book{lichnerowicz1967relativistic,
  title={Relativistic hydrodynamics and magnetohydrodynamics},
  author={Lichnerowicz, Andr{\'e} and Southwest Center for Advanced Studies and The Mathematical Physics Monagraph Series},
  volume={35},
  year={1967},
  publisher={WA Benjamin New York}
}


@article{chamel2013stability,
  title={Stability of super-Chandrasekhar magnetic white dwarfs},
  author={Chamel, Nicolas and Fantina, AF and Davis, PJ},
  journal={Physical Review D},
  volume={88},
  number={8},
  pages={081301},
  year={2013},
  publisher={APS}
}



@article{cardall2001effects,
  title={Effects of strong magnetic fields on neutron star structure},
  author={Cardall, Christian Y and Prakash, Madappa and Lattimer, James M},
  journal={The Astrophysical Journal},
  volume={554},
  number={1},
  pages={322},
  year={2001},
  publisher={IOP Publishing}
}

@article{paret2015maximum,
  title={Maximum mass of magnetic white dwarfs},
  author={Paret, D Manreza and Martinez, A Perez and Horvath, JE},
  journal={arXiv preprint arXiv:1501.04619},
  year={2015}
}



@article{Mereghetti:2010id,
      author         = "Mereghetti, S. and Tiengo, A. and Esposito, P. and La
                        Palombara, N. and Israel, G. L. and Stella, L.",
      title          = "{An ultra-massive fast-spinning white dwarf in a peculiar
                        binary system}",
      journal        = "Science",
      volume         = "325",
      year           = "2009",
      pages          = "1222",
      doi            = "10.1126/science.1176252",
      eprint         = "1003.0997",
      archivePrefix  = "arXiv",
      primaryClass   = "astro-ph.HE",
      SLACcitation   = "
}

@inproceedings{Malheiro:2015yda,
      author         = "Malheiro, M. and Coelho, J. G.",
      title          = "{MAGNETIC FIELDS OF SGRs/AXPs AS ROTATION-POWERED MASSIVE
                        WHITE DWARF PULSARS}",
      booktitle      = "{Proceedings, 13th Marcel Grossmann Meeting on Recent
                        Developments in Theoretical and Experimental General
                        Relativity, Astrophysics, and Relativistic Field Theories
                        (MG13)}",
      year           = "2015",
      pages          = "2462-2464",
      doi            = "10.1142/9789814623995_0470",
      SLACcitation   = "
}


@article{ostriker1969oscillations,
  title={On the Oscillations and Stability of Rotating Stellar Models. II. Rapidly Rotating White Dwarfs},
  author={Ostriker, Jeremiah P and Tassoul, Jean Louis},
  journal={The Astrophysical Journal},
  volume={155},
  pages={987},
  year={1969}
}

@article{ostriker1968rapidly1,
  title={Rapidly rotating stars. IV. Magnetic white dwarfs},
  author={Ostriker, Jeremiah P and Hartwick, FDA},
  journal={The Astrophysical Journal},
  volume={153},
  pages={797},
  year={1968}
}

@article{ostriker1968rapidly,
  title={Rapidly rotating stars. II. Massive white dwarfs},
  author={Ostriker, Jeremiah P and Bodenheimer, Peter},
  journal={The Astrophysical Journal},
  volume={151},
  pages={1089},
  year={1968}
}


@article{adam1986models,
  title={Models of magnetic white dwarfs},
  author={Adam, D},
  journal={Astronomy and Astrophysics},
  volume={160},
  pages={95--106},
  year={1986}
}

@article{suh2000mass,
  title={Mass-radius relation for magnetic white dwarfs},
  author={Suh, In-Saeng and Mathews, GJ},
  journal={The Astrophysical Journal},
  volume={530},
  number={2},
  pages={949},
  year={2000},
  publisher={IOP Publishing}
}

@book{chandrasekhar1939,
  title={An Introduction to the Study of Stellar Structure},
  author={Chandrasekhar, S.},
  year={1939},
  publisher={Chicago : Univ. Chicago Press}
}























\begin{thebibliography}{71}%
\makeatletter
\providecommand \@ifxundefined [1]{%
 \@ifx{#1\undefined}
}%
\providecommand \@ifnum [1]{%
 \ifnum #1\expandafter \@firstoftwo
 \else \expandafter \@secondoftwo
 \fi
}%
\providecommand \@ifx [1]{%
 \ifx #1\expandafter \@firstoftwo
 \else \expandafter \@secondoftwo
 \fi
}%
\providecommand \natexlab [1]{#1}%
\providecommand \enquote  [1]{``#1''}%
\providecommand \bibnamefont  [1]{#1}%
\providecommand \bibfnamefont [1]{#1}%
\providecommand \citenamefont [1]{#1}%
\providecommand \href@noop [0]{\@secondoftwo}%
\providecommand \href [0]{\begingroup \@sanitize@url \@href}%
\providecommand \@href[1]{\@@startlink{#1}\@@href}%
\providecommand \@@href[1]{\endgroup#1\@@endlink}%
\providecommand \@sanitize@url [0]{\catcode `\\12\catcode `\$12\catcode
  `\&12\catcode `\#12\catcode `\^12\catcode `\_12\catcode `\%12\relax}%
\providecommand \@@startlink[1]{}%
\providecommand \@@endlink[0]{}%
\providecommand \url  [0]{\begingroup\@sanitize@url \@url }%
\providecommand \@url [1]{\endgroup\@href {#1}{\urlprefix }}%
\providecommand \urlprefix  [0]{URL }%
\providecommand \Eprint [0]{\href }%
\providecommand \doibase [0]{http://dx.doi.org/}%
\providecommand \selectlanguage [0]{\@gobble}%
\providecommand \bibinfo  [0]{\@secondoftwo}%
\providecommand \bibfield  [0]{\@secondoftwo}%
\providecommand \translation [1]{[#1]}%
\providecommand \BibitemOpen [0]{}%
\providecommand \bibitemStop [0]{}%
\providecommand \bibitemNoStop [0]{.\EOS\space}%
\providecommand \EOS [0]{\spacefactor3000\relax}%
\providecommand \BibitemShut  [1]{\csname bibitem#1\endcsname}%
\let\auto@bib@innerbib\@empty
\bibitem [{\citenamefont {Shapiro}\ and\ \citenamefont
  {Teukolsky}(2008)}]{shapiro2008black}%
  \BibitemOpen
  \bibfield  {author} {\bibinfo {author} {\bibfnamefont {S.~L.}\ \bibnamefont
  {Shapiro}}\ and\ \bibinfo {author} {\bibfnamefont {S.~A.}\ \bibnamefont
  {Teukolsky}},\ }\href@noop {} {\emph {\bibinfo {title} {Black holes, white
  dwarfs and neutron stars: the physics of compact objects}}}\ (\bibinfo
  {publisher} {John Wiley \& Sons},\ \bibinfo {year} {2008})\BibitemShut
  {NoStop}%
\bibitem [{\citenamefont {Glendenning}(2012)}]{glendenning2012compact}%
  \BibitemOpen
  \bibfield  {author} {\bibinfo {author} {\bibfnamefont {N.~K.}\ \bibnamefont
  {Glendenning}},\ }\href@noop {} {\emph {\bibinfo {title} {Compact stars:
  Nuclear physics, particle physics and general relativity}}}\ (\bibinfo
  {publisher} {Springer Science \& Business Media},\ \bibinfo {year}
  {2012})\BibitemShut {NoStop}%
\bibitem [{\citenamefont {Fowler}(1926)}]{fowler1926dense}%
  \BibitemOpen
  \bibfield  {author} {\bibinfo {author} {\bibfnamefont {R.~H.}\ \bibnamefont
  {Fowler}},\ }\href@noop {} {\bibfield  {journal} {\bibinfo  {journal}
  {Monthly Notices of the Royal Astronomical Society}\ }\textbf {\bibinfo
  {volume} {87}},\ \bibinfo {pages} {114} (\bibinfo {year} {1926})}\BibitemShut
  {NoStop}%
\bibitem [{\citenamefont {Dirac}(1926)}]{dirac1926theory}%
  \BibitemOpen
  \bibfield  {author} {\bibinfo {author} {\bibfnamefont {P.~A.}\ \bibnamefont
  {Dirac}},\ }in\ \href@noop {} {\emph {\bibinfo {booktitle} {Proceedings of
  the Royal Society of London A: Mathematical, Physical and Engineering
  Sciences}}},\ Vol.\ \bibinfo {volume} {112}\ (\bibinfo {organization} {The
  Royal Society},\ \bibinfo {year} {1926})\ pp.\ \bibinfo {pages}
  {661--677}\BibitemShut {NoStop}%
\bibitem [{\citenamefont {Fermi}(1926)}]{fermi1926sulla}%
  \BibitemOpen
  \bibfield  {author} {\bibinfo {author} {\bibfnamefont {E.}~\bibnamefont
  {Fermi}},\ }\href@noop {} {\bibfield  {journal} {\bibinfo  {journal} {Rend.
  Lincei}\ }\textbf {\bibinfo {volume} {3}},\ \bibinfo {pages} {145} (\bibinfo
  {year} {1926})}\BibitemShut {NoStop}%
\bibitem [{\citenamefont {Chandrasekhar}(1931)}]{chandrasekhar1931maximum}%
  \BibitemOpen
  \bibfield  {author} {\bibinfo {author} {\bibfnamefont {S.}~\bibnamefont
  {Chandrasekhar}},\ }\href@noop {} {\bibfield  {journal} {\bibinfo  {journal}
  {The Astrophysical Journal}\ }\textbf {\bibinfo {volume} {74}},\ \bibinfo
  {pages} {81} (\bibinfo {year} {1931})}\BibitemShut {NoStop}%
\bibitem [{\citenamefont {Chandrasekhar}(1939)}]{chandrasekhar1939}%
  \BibitemOpen
  \bibfield  {author} {\bibinfo {author} {\bibfnamefont {S.}~\bibnamefont
  {Chandrasekhar}},\ }\href@noop {} {\emph {\bibinfo {title} {An Introduction
  to the Study of Stellar Structure}}}\ (\bibinfo  {publisher} {Chicago : Univ.
  Chicago Press},\ \bibinfo {year} {1939})\BibitemShut {NoStop}%
\bibitem [{\citenamefont {Terada}\ \emph {et~al.}(2008)\citenamefont {Terada},
  \citenamefont {Hayashi}, \citenamefont {Ishida}, \citenamefont {Mukai},
  \citenamefont {Dotani}, \citenamefont {Okada}, \citenamefont {Nakamura},
  \citenamefont {Naik}, \citenamefont {Bamba},\ and\ \citenamefont
  {Makishima}}]{Terada:2007br}%
  \BibitemOpen
  \bibfield  {author} {\bibinfo {author} {\bibfnamefont {Y.}~\bibnamefont
  {Terada}}, \bibinfo {author} {\bibfnamefont {T.}~\bibnamefont {Hayashi}},
  \bibinfo {author} {\bibfnamefont {M.}~\bibnamefont {Ishida}}, \bibinfo
  {author} {\bibfnamefont {K.}~\bibnamefont {Mukai}}, \bibinfo {author}
  {\bibfnamefont {T.~u.}\ \bibnamefont {Dotani}}, \bibinfo {author}
  {\bibfnamefont {S.}~\bibnamefont {Okada}}, \bibinfo {author} {\bibfnamefont
  {R.}~\bibnamefont {Nakamura}}, \bibinfo {author} {\bibfnamefont
  {S.}~\bibnamefont {Naik}}, \bibinfo {author} {\bibfnamefont {A.}~\bibnamefont
  {Bamba}}, \ and\ \bibinfo {author} {\bibfnamefont {K.}~\bibnamefont
  {Makishima}},\ }\href {\doibase 10.1093/pasj/60.2.387} {\bibfield  {journal}
  {\bibinfo  {journal} {Publ. Astron. Soc. Jap.}\ }\textbf {\bibinfo {volume}
  {60}},\ \bibinfo {pages} {387} (\bibinfo {year} {2008})},\ \Eprint
  {http://arxiv.org/abs/0711.2716} {arXiv:0711.2716 [astro-ph]} \BibitemShut
  {NoStop}%
\bibitem [{\citenamefont {Reimers}\ \emph {et~al.}(1996)\citenamefont
  {Reimers}, \citenamefont {Jordan}, \citenamefont {Koester}, \citenamefont
  {Bade}, \citenamefont {Kohler},\ and\ \citenamefont
  {Wisotzki}}]{Reimers:1995ia}%
  \BibitemOpen
  \bibfield  {author} {\bibinfo {author} {\bibfnamefont {D.}~\bibnamefont
  {Reimers}}, \bibinfo {author} {\bibfnamefont {S.}~\bibnamefont {Jordan}},
  \bibinfo {author} {\bibfnamefont {D.}~\bibnamefont {Koester}}, \bibinfo
  {author} {\bibfnamefont {N.}~\bibnamefont {Bade}}, \bibinfo {author}
  {\bibfnamefont {T.}~\bibnamefont {Kohler}}, \ and\ \bibinfo {author}
  {\bibfnamefont {L.}~\bibnamefont {Wisotzki}},\ }\href@noop {} {\bibfield
  {journal} {\bibinfo  {journal} {Astron. Astrophys.}\ }\textbf {\bibinfo
  {volume} {311}},\ \bibinfo {pages} {572} (\bibinfo {year} {1996})},\ \Eprint
  {http://arxiv.org/abs/astro-ph/9604104} {arXiv:astro-ph/9604104 [astro-ph]}
  \BibitemShut {NoStop}%
\bibitem [{\citenamefont {Schmidt}\ and\ \citenamefont
  {Smith}(1995)}]{Schmidt:1995eh}%
  \BibitemOpen
  \bibfield  {author} {\bibinfo {author} {\bibfnamefont {G.~D.}\ \bibnamefont
  {Schmidt}}\ and\ \bibinfo {author} {\bibfnamefont {P.~S.}\ \bibnamefont
  {Smith}},\ }\href {\doibase 10.1086/175962} {\bibfield  {journal} {\bibinfo
  {journal} {Astrophys. J.}\ }\textbf {\bibinfo {volume} {448}},\ \bibinfo
  {pages} {305} (\bibinfo {year} {1995})}\BibitemShut {NoStop}%
\bibitem [{\citenamefont {Kemp}\ \emph {et~al.}(1970)\citenamefont {Kemp},
  \citenamefont {Swedlund}, \citenamefont {Landstreet},\ and\ \citenamefont
  {Angel}}]{Kemp:1970zz}%
  \BibitemOpen
  \bibfield  {author} {\bibinfo {author} {\bibfnamefont {J.~C.}\ \bibnamefont
  {Kemp}}, \bibinfo {author} {\bibfnamefont {J.~B.}\ \bibnamefont {Swedlund}},
  \bibinfo {author} {\bibfnamefont {J.~D.}\ \bibnamefont {Landstreet}}, \ and\
  \bibinfo {author} {\bibfnamefont {J.~R.~P.}\ \bibnamefont {Angel}},\ }\href
  {\doibase 10.1086/180574} {\bibfield  {journal} {\bibinfo  {journal}
  {Astrophys. J.}\ }\textbf {\bibinfo {volume} {161}},\ \bibinfo {pages} {L77}
  (\bibinfo {year} {1970})}\BibitemShut {NoStop}%
\bibitem [{\citenamefont {Putney}(1995)}]{putney1995three}%
  \BibitemOpen
  \bibfield  {author} {\bibinfo {author} {\bibfnamefont {A.}~\bibnamefont
  {Putney}},\ }\href@noop {} {\bibfield  {journal} {\bibinfo  {journal} {The
  Astrophysical Journal Letters}\ }\textbf {\bibinfo {volume} {451}},\ \bibinfo
  {pages} {L67} (\bibinfo {year} {1995})}\BibitemShut {NoStop}%
\bibitem [{\citenamefont {Angel}(1978)}]{angel1978magnetic}%
  \BibitemOpen
  \bibfield  {author} {\bibinfo {author} {\bibfnamefont {J.}~\bibnamefont
  {Angel}},\ }\href@noop {} {\bibfield  {journal} {\bibinfo  {journal} {Annual
  Review of Astronomy and Astrophysics}\ }\textbf {\bibinfo {volume} {16}},\
  \bibinfo {pages} {487} (\bibinfo {year} {1978})}\BibitemShut {NoStop}%
\bibitem [{\citenamefont {Bera}\ and\ \citenamefont
  {Bhattacharya}(2014)}]{Bera:2014wja}%
  \BibitemOpen
  \bibfield  {author} {\bibinfo {author} {\bibfnamefont {P.}~\bibnamefont
  {Bera}}\ and\ \bibinfo {author} {\bibfnamefont {D.}~\bibnamefont
  {Bhattacharya}},\ }\href {\doibase 10.1093/mnras/stu2014} {\bibfield
  {journal} {\bibinfo  {journal} {Mon. Not. Roy. Astron. Soc.}\ }\textbf
  {\bibinfo {volume} {445}},\ \bibinfo {pages} {3951} (\bibinfo {year}
  {2014})},\ \Eprint {http://arxiv.org/abs/1405.2282} {arXiv:1405.2282
  [astro-ph.SR]} \BibitemShut {NoStop}%
\bibitem [{\citenamefont {Bocquet}\ \emph {et~al.}(1995)\citenamefont
  {Bocquet}, \citenamefont {Bonazzola}, \citenamefont {Gourgoulhon},\ and\
  \citenamefont {Novak}}]{Bocquet:1995je}%
  \BibitemOpen
  \bibfield  {author} {\bibinfo {author} {\bibfnamefont {M.}~\bibnamefont
  {Bocquet}}, \bibinfo {author} {\bibfnamefont {S.}~\bibnamefont {Bonazzola}},
  \bibinfo {author} {\bibfnamefont {E.}~\bibnamefont {Gourgoulhon}}, \ and\
  \bibinfo {author} {\bibfnamefont {J.}~\bibnamefont {Novak}},\ }\href@noop {}
  {\bibfield  {journal} {\bibinfo  {journal} {Astron. Astrophys.}\ }\textbf
  {\bibinfo {volume} {301}},\ \bibinfo {pages} {757} (\bibinfo {year}
  {1995})},\ \Eprint {http://arxiv.org/abs/gr-qc/9503044} {arXiv:gr-qc/9503044
  [gr-qc]} \BibitemShut {NoStop}%
\bibitem [{\citenamefont {Chatterjee}\ \emph {et~al.}(2015)\citenamefont
  {Chatterjee}, \citenamefont {Elghozi}, \citenamefont {Novak},\ and\
  \citenamefont {Oertel}}]{Chatterjee:2014qsa}%
  \BibitemOpen
  \bibfield  {author} {\bibinfo {author} {\bibfnamefont {D.}~\bibnamefont
  {Chatterjee}}, \bibinfo {author} {\bibfnamefont {T.}~\bibnamefont {Elghozi}},
  \bibinfo {author} {\bibfnamefont {J.}~\bibnamefont {Novak}}, \ and\ \bibinfo
  {author} {\bibfnamefont {M.}~\bibnamefont {Oertel}},\ }\href {\doibase
  10.1093/mnras/stu2706} {\bibfield  {journal} {\bibinfo  {journal} {Mon. Not.
  Roy. Astron. Soc.}\ }\textbf {\bibinfo {volume} {447}},\ \bibinfo {pages}
  {3785} (\bibinfo {year} {2015})},\ \Eprint {http://arxiv.org/abs/1410.6332}
  {arXiv:1410.6332 [astro-ph.HE]} \BibitemShut {NoStop}%
\bibitem [{\citenamefont {Cardall}\ \emph {et~al.}(2001)\citenamefont
  {Cardall}, \citenamefont {Prakash},\ and\ \citenamefont
  {Lattimer}}]{cardall2001effects}%
  \BibitemOpen
  \bibfield  {author} {\bibinfo {author} {\bibfnamefont {C.~Y.}\ \bibnamefont
  {Cardall}}, \bibinfo {author} {\bibfnamefont {M.}~\bibnamefont {Prakash}}, \
  and\ \bibinfo {author} {\bibfnamefont {J.~M.}\ \bibnamefont {Lattimer}},\
  }\href@noop {} {\bibfield  {journal} {\bibinfo  {journal} {The Astrophysical
  Journal}\ }\textbf {\bibinfo {volume} {554}},\ \bibinfo {pages} {322}
  (\bibinfo {year} {2001})}\BibitemShut {NoStop}%
\bibitem [{\citenamefont {Franzon}\ \emph {et~al.}(2015)\citenamefont
  {Franzon}, \citenamefont {Dexheimer},\ and\ \citenamefont
  {Schramm}}]{Franzon:2015sya}%
  \BibitemOpen
  \bibfield  {author} {\bibinfo {author} {\bibfnamefont {B.}~\bibnamefont
  {Franzon}}, \bibinfo {author} {\bibfnamefont {V.}~\bibnamefont {Dexheimer}},
  \ and\ \bibinfo {author} {\bibfnamefont {S.}~\bibnamefont {Schramm}},\
  }\href@noop {} {\  (\bibinfo {year} {2015})},\ \Eprint
  {http://arxiv.org/abs/1508.04431} {arXiv:1508.04431 [astro-ph.HE]}
  \BibitemShut {NoStop}%
\bibitem [{\citenamefont {Scalzo}\ \emph {et~al.}(2010)\citenamefont {Scalzo}
  \emph {et~al.}}]{Scalzo:2010xd}%
  \BibitemOpen
  \bibfield  {author} {\bibinfo {author} {\bibfnamefont {R.~A.}\ \bibnamefont
  {Scalzo}} \emph {et~al.},\ }\href {\doibase 10.1088/0004-637X/713/2/1073}
  {\bibfield  {journal} {\bibinfo  {journal} {Astrophys. J.}\ }\textbf
  {\bibinfo {volume} {713}},\ \bibinfo {pages} {1073} (\bibinfo {year}
  {2010})},\ \Eprint {http://arxiv.org/abs/1003.2217} {arXiv:1003.2217
  [astro-ph.CO]} \BibitemShut {NoStop}%
\bibitem [{\citenamefont {Howell}\ \emph {et~al.}(2006)\citenamefont {Howell}
  \emph {et~al.}}]{Howell:2006vn}%
  \BibitemOpen
  \bibfield  {author} {\bibinfo {author} {\bibfnamefont {D.~A.}\ \bibnamefont
  {Howell}} \emph {et~al.} (\bibinfo {collaboration} {SNLS}),\ }\href {\doibase
  10.1038/nature05103} {\bibfield  {journal} {\bibinfo  {journal} {Nature}\
  }\textbf {\bibinfo {volume} {443}},\ \bibinfo {pages} {308} (\bibinfo {year}
  {2006})},\ \Eprint {http://arxiv.org/abs/astro-ph/0609616}
  {arXiv:astro-ph/0609616 [astro-ph]} \BibitemShut {NoStop}%
\bibitem [{\citenamefont {Hicken}\ \emph {et~al.}(2007)\citenamefont {Hicken},
  \citenamefont {Garnavich}, \citenamefont {Prieto}, \citenamefont {Blondin},
  \citenamefont {DePoy}, \citenamefont {Kirshner},\ and\ \citenamefont
  {Parrent}}]{Hicken:2007ap}%
  \BibitemOpen
  \bibfield  {author} {\bibinfo {author} {\bibfnamefont {M.}~\bibnamefont
  {Hicken}}, \bibinfo {author} {\bibfnamefont {P.~M.}\ \bibnamefont
  {Garnavich}}, \bibinfo {author} {\bibfnamefont {J.~L.}\ \bibnamefont
  {Prieto}}, \bibinfo {author} {\bibfnamefont {S.}~\bibnamefont {Blondin}},
  \bibinfo {author} {\bibfnamefont {D.~L.}\ \bibnamefont {DePoy}}, \bibinfo
  {author} {\bibfnamefont {R.~P.}\ \bibnamefont {Kirshner}}, \ and\ \bibinfo
  {author} {\bibfnamefont {J.}~\bibnamefont {Parrent}},\ }\href {\doibase
  10.1086/523301} {\bibfield  {journal} {\bibinfo  {journal} {Astrophys. J.}\
  }\textbf {\bibinfo {volume} {669}},\ \bibinfo {pages} {L17} (\bibinfo {year}
  {2007})},\ \Eprint {http://arxiv.org/abs/0709.1501} {arXiv:0709.1501
  [astro-ph]} \BibitemShut {NoStop}%
\bibitem [{\citenamefont {Yamanaka}\ \emph {et~al.}(2009)\citenamefont
  {Yamanaka} \emph {et~al.}}]{Yamanaka:2009dp}%
  \BibitemOpen
  \bibfield  {author} {\bibinfo {author} {\bibfnamefont {M.}~\bibnamefont
  {Yamanaka}} \emph {et~al.},\ }\href {\doibase 10.1088/0004-637X/707/2/L118}
  {\bibfield  {journal} {\bibinfo  {journal} {Astrophys. J.}\ }\textbf
  {\bibinfo {volume} {707}},\ \bibinfo {pages} {L118} (\bibinfo {year}
  {2009})},\ \Eprint {http://arxiv.org/abs/0908.2059} {arXiv:0908.2059
  [astro-ph.HE]} \BibitemShut {NoStop}%
\bibitem [{\citenamefont {Taubenberger}\ \emph {et~al.}(2011)\citenamefont
  {Taubenberger}, \citenamefont {Benetti}, \citenamefont {Childress},
  \citenamefont {Pakmor}, \citenamefont {Hachinger}, \citenamefont {Mazzali},
  \citenamefont {Stanishev}, \citenamefont {Elias-Rosa}, \citenamefont
  {Agnoletto}, \citenamefont {Bufano} \emph {et~al.}}]{taubenberger2011high}%
  \BibitemOpen
  \bibfield  {author} {\bibinfo {author} {\bibfnamefont {S.}~\bibnamefont
  {Taubenberger}}, \bibinfo {author} {\bibfnamefont {S.}~\bibnamefont
  {Benetti}}, \bibinfo {author} {\bibfnamefont {M.}~\bibnamefont {Childress}},
  \bibinfo {author} {\bibfnamefont {R.}~\bibnamefont {Pakmor}}, \bibinfo
  {author} {\bibfnamefont {S.}~\bibnamefont {Hachinger}}, \bibinfo {author}
  {\bibfnamefont {P.}~\bibnamefont {Mazzali}}, \bibinfo {author} {\bibfnamefont
  {V.}~\bibnamefont {Stanishev}}, \bibinfo {author} {\bibfnamefont
  {N.}~\bibnamefont {Elias-Rosa}}, \bibinfo {author} {\bibfnamefont
  {I.}~\bibnamefont {Agnoletto}}, \bibinfo {author} {\bibfnamefont
  {F.}~\bibnamefont {Bufano}},  \emph {et~al.},\ }\href@noop {} {\bibfield
  {journal} {\bibinfo  {journal} {Monthly Notices of the Royal Astronomical
  Society}\ }\textbf {\bibinfo {volume} {412}},\ \bibinfo {pages} {2735}
  (\bibinfo {year} {2011})}\BibitemShut {NoStop}%
\bibitem [{\citenamefont {Moll}\ \emph {et~al.}(2014)\citenamefont {Moll},
  \citenamefont {Raskin}, \citenamefont {Kasen},\ and\ \citenamefont
  {Woosley}}]{Moll:2013mpa}%
  \BibitemOpen
  \bibfield  {author} {\bibinfo {author} {\bibfnamefont {R.}~\bibnamefont
  {Moll}}, \bibinfo {author} {\bibfnamefont {C.}~\bibnamefont {Raskin}},
  \bibinfo {author} {\bibfnamefont {D.}~\bibnamefont {Kasen}}, \ and\ \bibinfo
  {author} {\bibfnamefont {S.}~\bibnamefont {Woosley}},\ }\href {\doibase
  10.1088/0004-637X/785/2/105} {\bibfield  {journal} {\bibinfo  {journal}
  {Astrophys. J.}\ }\textbf {\bibinfo {volume} {785}},\ \bibinfo {pages} {105}
  (\bibinfo {year} {2014})},\ \Eprint {http://arxiv.org/abs/1311.5008}
  {arXiv:1311.5008 [astro-ph.HE]} \BibitemShut {NoStop}%
\bibitem [{\citenamefont {Das}\ and\ \citenamefont
  {Mukhopadhyay}(2012)}]{Das:2012ai}%
  \BibitemOpen
  \bibfield  {author} {\bibinfo {author} {\bibfnamefont {U.}~\bibnamefont
  {Das}}\ and\ \bibinfo {author} {\bibfnamefont {B.}~\bibnamefont
  {Mukhopadhyay}},\ }\href {\doibase 10.1103/PhysRevD.86.042001} {\bibfield
  {journal} {\bibinfo  {journal} {Phys. Rev.}\ }\textbf {\bibinfo {volume}
  {D86}},\ \bibinfo {pages} {042001} (\bibinfo {year} {2012})},\ \Eprint
  {http://arxiv.org/abs/1204.1262} {arXiv:1204.1262 [astro-ph.HE]} \BibitemShut
  {NoStop}%
\bibitem [{\citenamefont {Das}\ and\ \citenamefont
  {Mukhopadhyay}(2013)}]{Das:2013gd}%
  \BibitemOpen
  \bibfield  {author} {\bibinfo {author} {\bibfnamefont {U.}~\bibnamefont
  {Das}}\ and\ \bibinfo {author} {\bibfnamefont {B.}~\bibnamefont
  {Mukhopadhyay}},\ }\href {\doibase 10.1103/PhysRevLett.110.071102} {\bibfield
   {journal} {\bibinfo  {journal} {Phys.Rev.Lett.}\ }\textbf {\bibinfo {volume}
  {110}},\ \bibinfo {pages} {071102} (\bibinfo {year} {2013})},\ \Eprint
  {http://arxiv.org/abs/1301.5965} {arXiv:1301.5965 [astro-ph.SR]} \BibitemShut
  {NoStop}%
\bibitem [{\citenamefont {Das}\ and\ \citenamefont
  {Mukhopadhyay}(2014)}]{Das:2014ssa}%
  \BibitemOpen
  \bibfield  {author} {\bibinfo {author} {\bibfnamefont {U.}~\bibnamefont
  {Das}}\ and\ \bibinfo {author} {\bibfnamefont {B.}~\bibnamefont
  {Mukhopadhyay}},\ }\href {\doibase 10.1088/1475-7516/2014/06/050} {\bibfield
  {journal} {\bibinfo  {journal} {JCAP}\ }\textbf {\bibinfo {volume} {1406}},\
  \bibinfo {pages} {050} (\bibinfo {year} {2014})},\ \Eprint
  {http://arxiv.org/abs/1404.7627} {arXiv:1404.7627 [astro-ph.SR]} \BibitemShut
  {NoStop}%
\bibitem [{\citenamefont {Adam}(1986)}]{adam1986models}%
  \BibitemOpen
  \bibfield  {author} {\bibinfo {author} {\bibfnamefont {D.}~\bibnamefont
  {Adam}},\ }\href@noop {} {\bibfield  {journal} {\bibinfo  {journal}
  {Astronomy and Astrophysics}\ }\textbf {\bibinfo {volume} {160}},\ \bibinfo
  {pages} {95} (\bibinfo {year} {1986})}\BibitemShut {NoStop}%
\bibitem [{\citenamefont {Ostriker}\ and\ \citenamefont
  {Bodenheimer}(1968)}]{ostriker1968rapidly}%
  \BibitemOpen
  \bibfield  {author} {\bibinfo {author} {\bibfnamefont {J.~P.}\ \bibnamefont
  {Ostriker}}\ and\ \bibinfo {author} {\bibfnamefont {P.}~\bibnamefont
  {Bodenheimer}},\ }\href@noop {} {\bibfield  {journal} {\bibinfo  {journal}
  {The Astrophysical Journal}\ }\textbf {\bibinfo {volume} {151}},\ \bibinfo
  {pages} {1089} (\bibinfo {year} {1968})}\BibitemShut {NoStop}%
\bibitem [{\citenamefont {Ostriker}\ and\ \citenamefont
  {Hartwick}(1968)}]{ostriker1968rapidly1}%
  \BibitemOpen
  \bibfield  {author} {\bibinfo {author} {\bibfnamefont {J.~P.}\ \bibnamefont
  {Ostriker}}\ and\ \bibinfo {author} {\bibfnamefont {F.}~\bibnamefont
  {Hartwick}},\ }\href@noop {} {\bibfield  {journal} {\bibinfo  {journal} {The
  Astrophysical Journal}\ }\textbf {\bibinfo {volume} {153}},\ \bibinfo {pages}
  {797} (\bibinfo {year} {1968})}\BibitemShut {NoStop}%
\bibitem [{\citenamefont {Ostriker}\ and\ \citenamefont
  {Tassoul}(1969)}]{ostriker1969oscillations}%
  \BibitemOpen
  \bibfield  {author} {\bibinfo {author} {\bibfnamefont {J.~P.}\ \bibnamefont
  {Ostriker}}\ and\ \bibinfo {author} {\bibfnamefont {J.~L.}\ \bibnamefont
  {Tassoul}},\ }\href@noop {} {\bibfield  {journal} {\bibinfo  {journal} {The
  Astrophysical Journal}\ }\textbf {\bibinfo {volume} {155}},\ \bibinfo {pages}
  {987} (\bibinfo {year} {1969})}\BibitemShut {NoStop}%
\bibitem [{\citenamefont {Subramanian}\ and\ \citenamefont
  {Mukhopadhyay}(2015)}]{Subramanian:2015sza}%
  \BibitemOpen
  \bibfield  {author} {\bibinfo {author} {\bibfnamefont {S.}~\bibnamefont
  {Subramanian}}\ and\ \bibinfo {author} {\bibfnamefont {B.}~\bibnamefont
  {Mukhopadhyay}},\ }\href@noop {} {\  (\bibinfo {year} {2015})},\ \Eprint
  {http://arxiv.org/abs/1507.01606} {arXiv:1507.01606 [astro-ph.SR]}
  \BibitemShut {NoStop}%
\bibitem [{\citenamefont {Coelho}\ \emph {et~al.}(2014)\citenamefont {Coelho},
  \citenamefont {Marinho}, \citenamefont {Malheiro}, \citenamefont {Negreiros},
  \citenamefont {Rueda} \emph {et~al.}}]{Coelho:2013bba}%
  \BibitemOpen
  \bibfield  {author} {\bibinfo {author} {\bibfnamefont {J.}~\bibnamefont
  {Coelho}}, \bibinfo {author} {\bibfnamefont {R.}~\bibnamefont {Marinho}},
  \bibinfo {author} {\bibfnamefont {M.}~\bibnamefont {Malheiro}}, \bibinfo
  {author} {\bibfnamefont {R.}~\bibnamefont {Negreiros}}, \bibinfo {author}
  {\bibfnamefont {J.}~\bibnamefont {Rueda}},  \emph {et~al.},\ }\href {\doibase
  10.1088/0004-637X/794/1/86} {\bibfield  {journal} {\bibinfo  {journal}
  {Astrophys.J.}\ }\textbf {\bibinfo {volume} {794}},\ \bibinfo {pages} {86}
  (\bibinfo {year} {2014})},\ \Eprint {http://arxiv.org/abs/1306.4658}
  {arXiv:1306.4658 [astro-ph.SR]} \BibitemShut {NoStop}%
\bibitem [{\citenamefont {Chamel}\ \emph {et~al.}(2013)\citenamefont {Chamel},
  \citenamefont {Fantina},\ and\ \citenamefont {Davis}}]{PhysRevD.88.081301}%
  \BibitemOpen
  \bibfield  {author} {\bibinfo {author} {\bibfnamefont {N.}~\bibnamefont
  {Chamel}}, \bibinfo {author} {\bibfnamefont {A.~F.}\ \bibnamefont {Fantina}},
  \ and\ \bibinfo {author} {\bibfnamefont {P.~J.}\ \bibnamefont {Davis}},\
  }\href {\doibase 10.1103/PhysRevD.88.081301} {\bibfield  {journal} {\bibinfo
  {journal} {Phys. Rev. D}\ }\textbf {\bibinfo {volume} {88}},\ \bibinfo
  {pages} {081301} (\bibinfo {year} {2013})}\BibitemShut {NoStop}%
\bibitem [{\citenamefont {Nityananda}\ and\ \citenamefont
  {Konar}(2015)}]{PhysRevD.91.028301}%
  \BibitemOpen
  \bibfield  {author} {\bibinfo {author} {\bibfnamefont {R.}~\bibnamefont
  {Nityananda}}\ and\ \bibinfo {author} {\bibfnamefont {S.}~\bibnamefont
  {Konar}},\ }\href {\doibase 10.1103/PhysRevD.91.028301} {\bibfield  {journal}
  {\bibinfo  {journal} {Phys. Rev. D}\ }\textbf {\bibinfo {volume} {91}},\
  \bibinfo {pages} {028301} (\bibinfo {year} {2015})}\BibitemShut {NoStop}%
\bibitem [{\citenamefont {Bonazzola}\ \emph {et~al.}(1993)\citenamefont
  {Bonazzola}, \citenamefont {Gourgoulhon}, \citenamefont {Salgado},\ and\
  \citenamefont {Marck}}]{Bonazzola:1993zz}%
  \BibitemOpen
  \bibfield  {author} {\bibinfo {author} {\bibfnamefont {S.}~\bibnamefont
  {Bonazzola}}, \bibinfo {author} {\bibfnamefont {E.}~\bibnamefont
  {Gourgoulhon}}, \bibinfo {author} {\bibfnamefont {M.}~\bibnamefont
  {Salgado}}, \ and\ \bibinfo {author} {\bibfnamefont {J.~A. t. A. r. r. b. A.
  n. n. a. f. e.~s.}\ \bibnamefont {Marck}},\ }\href@noop {} {\bibfield
  {journal} {\bibinfo  {journal} {Astron. Astrophys.}\ }\textbf {\bibinfo
  {volume} {278}},\ \bibinfo {pages} {421} (\bibinfo {year}
  {1993})}\BibitemShut {NoStop}%
\bibitem [{\citenamefont {Gourgoulhon}(2012)}]{gourgoulhon20123+}%
  \BibitemOpen
  \bibfield  {author} {\bibinfo {author} {\bibfnamefont {E.}~\bibnamefont
  {Gourgoulhon}},\ }\href@noop {} {\emph {\bibinfo {title} {3+ 1 formalism in
  general relativity: bases of numerical relativity}}},\ Vol.\ \bibinfo
  {volume} {846}\ (\bibinfo  {publisher} {Springer Science \& Business Media},\
  \bibinfo {year} {2012})\BibitemShut {NoStop}%
\bibitem [{\citenamefont {Lichnerowicz}\ \emph {et~al.}(1967)\citenamefont
  {Lichnerowicz}, \citenamefont {for Advanced~Studies},\ and\ \citenamefont
  {Series}}]{lichnerowicz1967relativistic}%
  \BibitemOpen
  \bibfield  {author} {\bibinfo {author} {\bibfnamefont {A.}~\bibnamefont
  {Lichnerowicz}}, \bibinfo {author} {\bibfnamefont {S.~C.}\ \bibnamefont {for
  Advanced~Studies}}, \ and\ \bibinfo {author} {\bibfnamefont {T.~M. P.~M.}\
  \bibnamefont {Series}},\ }\href@noop {} {\emph {\bibinfo {title}
  {Relativistic hydrodynamics and magnetohydrodynamics}}},\ Vol.~\bibinfo
  {volume} {35}\ (\bibinfo  {publisher} {WA Benjamin New York},\ \bibinfo
  {year} {1967})\BibitemShut {NoStop}%
\bibitem [{\citenamefont {Suh}\ and\ \citenamefont
  {Mathews}(2000)}]{suh2000mass}%
  \BibitemOpen
  \bibfield  {author} {\bibinfo {author} {\bibfnamefont {I.-S.}\ \bibnamefont
  {Suh}}\ and\ \bibinfo {author} {\bibfnamefont {G.}~\bibnamefont {Mathews}},\
  }\href@noop {} {\bibfield  {journal} {\bibinfo  {journal} {The Astrophysical
  Journal}\ }\textbf {\bibinfo {volume} {530}},\ \bibinfo {pages} {949}
  (\bibinfo {year} {2000})}\BibitemShut {NoStop}%
\bibitem [{\citenamefont {Chamel}\ and\ \citenamefont
  {Fantina}(2015)}]{PhysRevD.92.023008}%
  \BibitemOpen
  \bibfield  {author} {\bibinfo {author} {\bibfnamefont {N.}~\bibnamefont
  {Chamel}}\ and\ \bibinfo {author} {\bibfnamefont {A.~F.}\ \bibnamefont
  {Fantina}},\ }\href {\doibase 10.1103/PhysRevD.92.023008} {\bibfield
  {journal} {\bibinfo  {journal} {Phys. Rev. D}\ }\textbf {\bibinfo {volume}
  {92}},\ \bibinfo {pages} {023008} (\bibinfo {year} {2015})}\BibitemShut
  {NoStop}%
\bibitem [{\citenamefont {Paret}\ \emph {et~al.}(2015)\citenamefont {Paret},
  \citenamefont {Martinez},\ and\ \citenamefont {Horvath}}]{paret2015maximum}%
  \BibitemOpen
  \bibfield  {author} {\bibinfo {author} {\bibfnamefont {D.~M.}\ \bibnamefont
  {Paret}}, \bibinfo {author} {\bibfnamefont {A.~P.}\ \bibnamefont {Martinez}},
  \ and\ \bibinfo {author} {\bibfnamefont {J.}~\bibnamefont {Horvath}},\
  }\href@noop {} {\bibfield  {journal} {\bibinfo  {journal} {arXiv preprint
  arXiv:1501.04619}\ } (\bibinfo {year} {2015})}\BibitemShut {NoStop}%
\bibitem [{\citenamefont {Mereghetti}\ \emph {et~al.}(2009)\citenamefont
  {Mereghetti}, \citenamefont {Tiengo}, \citenamefont {Esposito}, \citenamefont
  {La~Palombara}, \citenamefont {Israel},\ and\ \citenamefont
  {Stella}}]{Mereghetti:2010id}%
  \BibitemOpen
  \bibfield  {author} {\bibinfo {author} {\bibfnamefont {S.}~\bibnamefont
  {Mereghetti}}, \bibinfo {author} {\bibfnamefont {A.}~\bibnamefont {Tiengo}},
  \bibinfo {author} {\bibfnamefont {P.}~\bibnamefont {Esposito}}, \bibinfo
  {author} {\bibfnamefont {N.}~\bibnamefont {La~Palombara}}, \bibinfo {author}
  {\bibfnamefont {G.~L.}\ \bibnamefont {Israel}}, \ and\ \bibinfo {author}
  {\bibfnamefont {L.}~\bibnamefont {Stella}},\ }\href {\doibase
  10.1126/science.1176252} {\bibfield  {journal} {\bibinfo  {journal}
  {Science}\ }\textbf {\bibinfo {volume} {325}},\ \bibinfo {pages} {1222}
  (\bibinfo {year} {2009})},\ \Eprint {http://arxiv.org/abs/1003.0997}
  {arXiv:1003.0997 [astro-ph.HE]} \BibitemShut {NoStop}%
\bibitem [{\citenamefont {Duncan}\ and\ \citenamefont
  {Thompson}(1992)}]{Duncan:1992hi}%
  \BibitemOpen
  \bibfield  {author} {\bibinfo {author} {\bibfnamefont {R.~C.}\ \bibnamefont
  {Duncan}}\ and\ \bibinfo {author} {\bibfnamefont {C.}~\bibnamefont
  {Thompson}},\ }\href {\doibase 10.1086/186413} {\bibfield  {journal}
  {\bibinfo  {journal} {Astrophys. J.}\ }\textbf {\bibinfo {volume} {392}},\
  \bibinfo {pages} {L9} (\bibinfo {year} {1992})}\BibitemShut {NoStop}%
\bibitem [{\citenamefont {Thompson}\ and\ \citenamefont
  {Duncan}(1993)}]{Thompson:1993hn}%
  \BibitemOpen
  \bibfield  {author} {\bibinfo {author} {\bibfnamefont {C.}~\bibnamefont
  {Thompson}}\ and\ \bibinfo {author} {\bibfnamefont {R.~C.}\ \bibnamefont
  {Duncan}},\ }\href {\doibase 10.1086/172580} {\bibfield  {journal} {\bibinfo
  {journal} {Astrophys. J.}\ }\textbf {\bibinfo {volume} {408}},\ \bibinfo
  {pages} {194} (\bibinfo {year} {1993})}\BibitemShut {NoStop}%
\bibitem [{\citenamefont {Malheiro}\ and\ \citenamefont
  {Coelho}(2015)}]{Malheiro:2015yda}%
  \BibitemOpen
  \bibfield  {author} {\bibinfo {author} {\bibfnamefont {M.}~\bibnamefont
  {Malheiro}}\ and\ \bibinfo {author} {\bibfnamefont {J.~G.}\ \bibnamefont
  {Coelho}},\ }in\ \href {\doibase 10.1142/9789814623995_0470} {\emph {\bibinfo
  {booktitle} {{Proceedings, 13th Marcel Grossmann Meeting on Recent
  Developments in Theoretical and Experimental General Relativity,
  Astrophysics, and Relativistic Field Theories (MG13)}}}}\ (\bibinfo {year}
  {2015})\ pp.\ \bibinfo {pages} {2462--2464}\BibitemShut {NoStop}%
\bibitem [{\citenamefont {Krishan}\ and\ \citenamefont
  {Kushwaha}(1963)}]{krishan1963limiting}%
  \BibitemOpen
  \bibfield  {author} {\bibinfo {author} {\bibfnamefont {S.}~\bibnamefont
  {Krishan}}\ and\ \bibinfo {author} {\bibfnamefont {R.}~\bibnamefont
  {Kushwaha}},\ }\href@noop {} {\bibfield  {journal} {\bibinfo  {journal}
  {Publications of the Astronomical Society of Japan}\ }\textbf {\bibinfo
  {volume} {15}},\ \bibinfo {pages} {253} (\bibinfo {year} {1963})}\BibitemShut
  {NoStop}%
\bibitem [{\citenamefont {Anand}(1965)}]{anand1965chandrasekhar}%
  \BibitemOpen
  \bibfield  {author} {\bibinfo {author} {\bibfnamefont {S.}~\bibnamefont
  {Anand}},\ }\href@noop {} {\bibfield  {journal} {\bibinfo  {journal}
  {Proceedings of the National Academy of Sciences of the United States of
  America}\ }\textbf {\bibinfo {volume} {54}},\ \bibinfo {pages} {23} (\bibinfo
  {year} {1965})}\BibitemShut {NoStop}%
\bibitem [{\citenamefont {James}(1964)}]{james1964structure}%
  \BibitemOpen
  \bibfield  {author} {\bibinfo {author} {\bibfnamefont {R.}~\bibnamefont
  {James}},\ }\href@noop {} {\bibfield  {journal} {\bibinfo  {journal} {The
  Astrophysical Journal}\ }\textbf {\bibinfo {volume} {140}},\ \bibinfo {pages}
  {552} (\bibinfo {year} {1964})}\BibitemShut {NoStop}%
\bibitem [{\citenamefont {Roxburgh}\ and\ \citenamefont
  {Durney}(1966)}]{roxburgh1966structure}%
  \BibitemOpen
  \bibfield  {author} {\bibinfo {author} {\bibfnamefont {I.}~\bibnamefont
  {Roxburgh}}\ and\ \bibinfo {author} {\bibfnamefont {B.}~\bibnamefont
  {Durney}},\ }\href@noop {} {\bibfield  {journal} {\bibinfo  {journal}
  {Zeitschrift fur Astrophysik}\ }\textbf {\bibinfo {volume} {64}},\ \bibinfo
  {pages} {504} (\bibinfo {year} {1966})}\BibitemShut {NoStop}%
\bibitem [{\citenamefont {Monaghan}(1966)}]{monaghan1966structure}%
  \BibitemOpen
  \bibfield  {author} {\bibinfo {author} {\bibfnamefont {J.}~\bibnamefont
  {Monaghan}},\ }\href@noop {} {\bibfield  {journal} {\bibinfo  {journal}
  {Monthly Notices of the Royal Astronomical Society}\ }\textbf {\bibinfo
  {volume} {132}},\ \bibinfo {pages} {305} (\bibinfo {year}
  {1966})}\BibitemShut {NoStop}%
\bibitem [{\citenamefont {Geroyannis}\ and\ \citenamefont
  {Hadjopoulos}(1989)}]{geroyannis1989models}%
  \BibitemOpen
  \bibfield  {author} {\bibinfo {author} {\bibfnamefont {V.}~\bibnamefont
  {Geroyannis}}\ and\ \bibinfo {author} {\bibfnamefont {A.}~\bibnamefont
  {Hadjopoulos}},\ }\href@noop {} {\bibfield  {journal} {\bibinfo  {journal}
  {The Astrophysical Journal Supplement Series}\ }\textbf {\bibinfo {volume}
  {70}},\ \bibinfo {pages} {661} (\bibinfo {year} {1989})}\BibitemShut
  {NoStop}%
\bibitem [{\citenamefont {Arutyunyan}\ \emph {et~al.}(1971)\citenamefont
  {Arutyunyan}, \citenamefont {Sedrakyan},\ and\ \citenamefont
  {Chubaryan}}]{arutyunyan1971rotating}%
  \BibitemOpen
  \bibfield  {author} {\bibinfo {author} {\bibfnamefont {G.}~\bibnamefont
  {Arutyunyan}}, \bibinfo {author} {\bibfnamefont {D.}~\bibnamefont
  {Sedrakyan}}, \ and\ \bibinfo {author} {\bibfnamefont {{\'E}.}~\bibnamefont
  {Chubaryan}},\ }\href@noop {} {\bibfield  {journal} {\bibinfo  {journal}
  {Soviet Astronomy}\ }\textbf {\bibinfo {volume} {15}},\ \bibinfo {pages}
  {390} (\bibinfo {year} {1971})}\BibitemShut {NoStop}%
\bibitem [{\citenamefont {Boshkayev}\ \emph {et~al.}(2013)\citenamefont
  {Boshkayev}, \citenamefont {Rueda}, \citenamefont {Ruffini},\ and\
  \citenamefont {Siutsou}}]{boshkayev2013general}%
  \BibitemOpen
  \bibfield  {author} {\bibinfo {author} {\bibfnamefont {K.}~\bibnamefont
  {Boshkayev}}, \bibinfo {author} {\bibfnamefont {J.~A.}\ \bibnamefont
  {Rueda}}, \bibinfo {author} {\bibfnamefont {R.}~\bibnamefont {Ruffini}}, \
  and\ \bibinfo {author} {\bibfnamefont {I.}~\bibnamefont {Siutsou}},\
  }\href@noop {} {\bibfield  {journal} {\bibinfo  {journal} {The Astrophysical
  Journal}\ }\textbf {\bibinfo {volume} {762}},\ \bibinfo {pages} {117}
  (\bibinfo {year} {2013})}\BibitemShut {NoStop}%
\bibitem [{\citenamefont {Hartle}(1967)}]{hartle1967slowly}%
  \BibitemOpen
  \bibfield  {author} {\bibinfo {author} {\bibfnamefont {J.~B.}\ \bibnamefont
  {Hartle}},\ }\href@noop {} {\bibfield  {journal} {\bibinfo  {journal} {The
  Astrophysical Journal}\ }\textbf {\bibinfo {volume} {150}},\ \bibinfo {pages}
  {1005} (\bibinfo {year} {1967})}\BibitemShut {NoStop}%
\bibitem [{\citenamefont {Markey}\ and\ \citenamefont
  {Tayler}(1973)}]{markey1973adiabatic}%
  \BibitemOpen
  \bibfield  {author} {\bibinfo {author} {\bibfnamefont {P.}~\bibnamefont
  {Markey}}\ and\ \bibinfo {author} {\bibfnamefont {R.}~\bibnamefont
  {Tayler}},\ }\href@noop {} {\bibfield  {journal} {\bibinfo  {journal}
  {Monthly Notices of the Royal Astronomical Society}\ }\textbf {\bibinfo
  {volume} {163}},\ \bibinfo {pages} {77} (\bibinfo {year} {1973})}\BibitemShut
  {NoStop}%
\bibitem [{\citenamefont {Tayler}(1973)}]{tayler1973adiabatic}%
  \BibitemOpen
  \bibfield  {author} {\bibinfo {author} {\bibfnamefont {R.}~\bibnamefont
  {Tayler}},\ }\href@noop {} {\bibfield  {journal} {\bibinfo  {journal}
  {Monthly Notices of the Royal Astronomical Society}\ }\textbf {\bibinfo
  {volume} {161}},\ \bibinfo {pages} {365} (\bibinfo {year}
  {1973})}\BibitemShut {NoStop}%
\bibitem [{\citenamefont {Wright}(1973)}]{wright1973pinch}%
  \BibitemOpen
  \bibfield  {author} {\bibinfo {author} {\bibfnamefont {G.}~\bibnamefont
  {Wright}},\ }\href@noop {} {\bibfield  {journal} {\bibinfo  {journal}
  {Monthly Notices of the Royal Astronomical Society}\ }\textbf {\bibinfo
  {volume} {162}},\ \bibinfo {pages} {339} (\bibinfo {year}
  {1973})}\BibitemShut {NoStop}%
\bibitem [{\citenamefont {Flowers}\ and\ \citenamefont
  {Ruderman}(1977)}]{flowers1977evolution}%
  \BibitemOpen
  \bibfield  {author} {\bibinfo {author} {\bibfnamefont {E.}~\bibnamefont
  {Flowers}}\ and\ \bibinfo {author} {\bibfnamefont {M.~A.}\ \bibnamefont
  {Ruderman}},\ }\href@noop {} {\bibfield  {journal} {\bibinfo  {journal} {The
  Astrophysical Journal}\ }\textbf {\bibinfo {volume} {215}},\ \bibinfo {pages}
  {302} (\bibinfo {year} {1977})}\BibitemShut {NoStop}%
\bibitem [{\citenamefont {Lander}\ and\ \citenamefont
  {Jones}(2012)}]{lander2012there}%
  \BibitemOpen
  \bibfield  {author} {\bibinfo {author} {\bibfnamefont {S.~K.}\ \bibnamefont
  {Lander}}\ and\ \bibinfo {author} {\bibfnamefont {D.}~\bibnamefont {Jones}},\
  }\href@noop {} {\bibfield  {journal} {\bibinfo  {journal} {Monthly Notices of
  the Royal Astronomical Society}\ }\textbf {\bibinfo {volume} {424}},\
  \bibinfo {pages} {482} (\bibinfo {year} {2012})}\BibitemShut {NoStop}%
\bibitem [{\citenamefont {Braithwaite}(2006)}]{braithwaite2006stability}%
  \BibitemOpen
  \bibfield  {author} {\bibinfo {author} {\bibfnamefont {J.}~\bibnamefont
  {Braithwaite}},\ }\href@noop {} {\bibfield  {journal} {\bibinfo  {journal}
  {Astronomy \& Astrophysics}\ }\textbf {\bibinfo {volume} {453}},\ \bibinfo
  {pages} {687} (\bibinfo {year} {2006})}\BibitemShut {NoStop}%
\bibitem [{\citenamefont {Braithwaite}\ and\ \citenamefont
  {Nordlund}(2006)}]{braithwaite2006stable}%
  \BibitemOpen
  \bibfield  {author} {\bibinfo {author} {\bibfnamefont {J.}~\bibnamefont
  {Braithwaite}}\ and\ \bibinfo {author} {\bibfnamefont {{\AA}.}~\bibnamefont
  {Nordlund}},\ }\href@noop {} {\bibfield  {journal} {\bibinfo  {journal}
  {Astronomy \& Astrophysics}\ }\textbf {\bibinfo {volume} {450}},\ \bibinfo
  {pages} {1077} (\bibinfo {year} {2006})}\BibitemShut {NoStop}%
\bibitem [{\citenamefont {Braithwaite}(2007)}]{braithwaite2007stability}%
  \BibitemOpen
  \bibfield  {author} {\bibinfo {author} {\bibfnamefont {J.}~\bibnamefont
  {Braithwaite}},\ }\href@noop {} {\bibfield  {journal} {\bibinfo  {journal}
  {Astronomy \& Astrophysics}\ }\textbf {\bibinfo {volume} {469}},\ \bibinfo
  {pages} {275} (\bibinfo {year} {2007})}\BibitemShut {NoStop}%
\bibitem [{\citenamefont {Ciolfi}\ and\ \citenamefont
  {Rezzolla}(2013)}]{ciolfi2013twisted}%
  \BibitemOpen
  \bibfield  {author} {\bibinfo {author} {\bibfnamefont {R.}~\bibnamefont
  {Ciolfi}}\ and\ \bibinfo {author} {\bibfnamefont {L.}~\bibnamefont
  {Rezzolla}},\ }\href@noop {} {\bibfield  {journal} {\bibinfo  {journal}
  {Monthly Notices of the Royal Astronomical Society: Letters}\ }\textbf
  {\bibinfo {volume} {435}},\ \bibinfo {pages} {L43} (\bibinfo {year}
  {2013})}\BibitemShut {NoStop}%
\bibitem [{\citenamefont {Lasky}\ \emph {et~al.}(2011)\citenamefont {Lasky},
  \citenamefont {Zink}, \citenamefont {Kokkotas},\ and\ \citenamefont
  {Glampedakis}}]{lasky2011hydromagnetic}%
  \BibitemOpen
  \bibfield  {author} {\bibinfo {author} {\bibfnamefont {P.~D.}\ \bibnamefont
  {Lasky}}, \bibinfo {author} {\bibfnamefont {B.}~\bibnamefont {Zink}},
  \bibinfo {author} {\bibfnamefont {K.~D.}\ \bibnamefont {Kokkotas}}, \ and\
  \bibinfo {author} {\bibfnamefont {K.}~\bibnamefont {Glampedakis}},\
  }\href@noop {} {\bibfield  {journal} {\bibinfo  {journal} {The Astrophysical
  Journal Letters}\ }\textbf {\bibinfo {volume} {735}},\ \bibinfo {pages} {L20}
  (\bibinfo {year} {2011})}\BibitemShut {NoStop}%
\bibitem [{\citenamefont {Marchant}\ \emph {et~al.}(2011)\citenamefont
  {Marchant}, \citenamefont {Reisenegger},\ and\ \citenamefont
  {Akg{\"u}n}}]{marchant2011revisiting}%
  \BibitemOpen
  \bibfield  {author} {\bibinfo {author} {\bibfnamefont {P.}~\bibnamefont
  {Marchant}}, \bibinfo {author} {\bibfnamefont {A.}~\bibnamefont
  {Reisenegger}}, \ and\ \bibinfo {author} {\bibfnamefont {T.}~\bibnamefont
  {Akg{\"u}n}},\ }\href@noop {} {\bibfield  {journal} {\bibinfo  {journal}
  {Monthly Notices of the Royal Astronomical Society}\ }\textbf {\bibinfo
  {volume} {415}},\ \bibinfo {pages} {2426} (\bibinfo {year}
  {2011})}\BibitemShut {NoStop}%
\bibitem [{\citenamefont {Mitchell}\ \emph {et~al.}(2015)\citenamefont
  {Mitchell}, \citenamefont {Braithwaite}, \citenamefont {Reisenegger},
  \citenamefont {Spruit}, \citenamefont {Valdivia},\ and\ \citenamefont
  {Langer}}]{mitchell2015instability}%
  \BibitemOpen
  \bibfield  {author} {\bibinfo {author} {\bibfnamefont {J.}~\bibnamefont
  {Mitchell}}, \bibinfo {author} {\bibfnamefont {J.}~\bibnamefont
  {Braithwaite}}, \bibinfo {author} {\bibfnamefont {A.}~\bibnamefont
  {Reisenegger}}, \bibinfo {author} {\bibfnamefont {H.}~\bibnamefont {Spruit}},
  \bibinfo {author} {\bibfnamefont {J.}~\bibnamefont {Valdivia}}, \ and\
  \bibinfo {author} {\bibfnamefont {N.}~\bibnamefont {Langer}},\ }\href@noop {}
  {\bibfield  {journal} {\bibinfo  {journal} {Monthly Notices of the Royal
  Astronomical Society}\ }\textbf {\bibinfo {volume} {447}},\ \bibinfo {pages}
  {1213} (\bibinfo {year} {2015})}\BibitemShut {NoStop}%
\bibitem [{\citenamefont {Armaza}\ \emph {et~al.}(2015)\citenamefont {Armaza},
  \citenamefont {Reisenegger},\ and\ \citenamefont
  {Valdivia}}]{armaza2015magnetic}%
  \BibitemOpen
  \bibfield  {author} {\bibinfo {author} {\bibfnamefont {C.}~\bibnamefont
  {Armaza}}, \bibinfo {author} {\bibfnamefont {A.}~\bibnamefont {Reisenegger}},
  \ and\ \bibinfo {author} {\bibfnamefont {J.~A.}\ \bibnamefont {Valdivia}},\
  }\href@noop {} {\bibfield  {journal} {\bibinfo  {journal} {The Astrophysical
  Journal}\ }\textbf {\bibinfo {volume} {802}},\ \bibinfo {pages} {121}
  (\bibinfo {year} {2015})}\BibitemShut {NoStop}%
\bibitem [{\citenamefont {Prendergast}(1956)}]{prendergast1956equilibrium}%
  \BibitemOpen
  \bibfield  {author} {\bibinfo {author} {\bibfnamefont {K.~H.}\ \bibnamefont
  {Prendergast}},\ }\href@noop {} {\bibfield  {journal} {\bibinfo  {journal}
  {The Astrophysical Journal}\ }\textbf {\bibinfo {volume} {123}},\ \bibinfo
  {pages} {498} (\bibinfo {year} {1956})}\BibitemShut {NoStop}%
\bibitem [{\citenamefont {Braithwaite}\ and\ \citenamefont
  {Spruit}(2004)}]{braithwaite2004fossil}%
  \BibitemOpen
  \bibfield  {author} {\bibinfo {author} {\bibfnamefont {J.}~\bibnamefont
  {Braithwaite}}\ and\ \bibinfo {author} {\bibfnamefont {H.~C.}\ \bibnamefont
  {Spruit}},\ }\href@noop {} {\bibfield  {journal} {\bibinfo  {journal}
  {Nature}\ }\textbf {\bibinfo {volume} {431}},\ \bibinfo {pages} {819}
  (\bibinfo {year} {2004})}\BibitemShut {NoStop}%
\bibitem [{\citenamefont {Akg{\"u}n}\ \emph {et~al.}(2013)\citenamefont
  {Akg{\"u}n}, \citenamefont {Reisenegger}, \citenamefont {Mastrano},\ and\
  \citenamefont {Marchant}}]{akgun2013stability}%
  \BibitemOpen
  \bibfield  {author} {\bibinfo {author} {\bibfnamefont {T.}~\bibnamefont
  {Akg{\"u}n}}, \bibinfo {author} {\bibfnamefont {A.}~\bibnamefont
  {Reisenegger}}, \bibinfo {author} {\bibfnamefont {A.}~\bibnamefont
  {Mastrano}}, \ and\ \bibinfo {author} {\bibfnamefont {P.}~\bibnamefont
  {Marchant}},\ }\href@noop {} {\bibfield  {journal} {\bibinfo  {journal}
  {Monthly Notices of the Royal Astronomical Society}\ }\textbf {\bibinfo
  {volume} {433}},\ \bibinfo {pages} {2445} (\bibinfo {year}
  {2013})}\BibitemShut {NoStop}%
\bibitem [{\citenamefont {Goldreich}\ and\ \citenamefont
  {Reisenegger}(1992)}]{goldreich1992magnetic}%
  \BibitemOpen
  \bibfield  {author} {\bibinfo {author} {\bibfnamefont {P.}~\bibnamefont
  {Goldreich}}\ and\ \bibinfo {author} {\bibfnamefont {A.}~\bibnamefont
  {Reisenegger}},\ }\href@noop {} {\bibfield  {journal} {\bibinfo  {journal}
  {The Astrophysical Journal}\ }\textbf {\bibinfo {volume} {395}},\ \bibinfo
  {pages} {250} (\bibinfo {year} {1992})}\BibitemShut {NoStop}%
\end{thebibliography}%

\end{document}